\documentclass[journal]{IEEEtran}
\usepackage{graphicx}
\usepackage{amsmath}
\DeclareGraphicsExtensions{.pdf,.jpeg,.png}
\ifCLASSOPTIONcompsoc
  \usepackage[caption=false, font=footnotesize, labelfont=sf, textfont=sf, labelformat=simple]{subfig}
\else
  \usepackage[caption=false, font=footnotesize, labelformat=simple]{subfig}
\fi

\usepackage{adjustbox}
\usepackage{xspace}
\newcommand{\BfPara}[1]{{\noindent\bf#1.}\xspace}
\usepackage{xcolor}
\PassOptionsToPackage{hyphens}{url}\usepackage{hyperref}
\hypersetup{colorlinks=true, urlcolor=blue, citecolor=blue, linkcolor=blue, pdfborder={0 0 0},}
\usepackage{tabularx}
\def\Snospace~{\S{}}

\usepackage{adjustbox}
\newcolumntype{R}[2]{%
    >{\adjustbox{angle=#1,lap=\width-(#2)}\bgroup}%
    r%
    <{\egroup}%
}
\newcommand*\rot{\multicolumn{1}{R{90}{.8em}}}
\usepackage{pifont}
\newcommand{\checkmark}{\ding{52}}

\usepackage{amsfonts}
\usepackage{multirow}

\usepackage[linesnumbered,ruled,vlined]{algorithm2e}

\usepackage[noabbrev,nameinlink]{cleveref}
\crefname{equation}{}{}
\IEEEaftertitletext{\vspace{-1\baselineskip}}
\begin{document}
\title{X-CANIDS: Signal-Aware Explainable Intrusion Detection System for Controller Area Network-Based In-Vehicle Network}
\author{Seonghoon Jeong,~\IEEEmembership{Member,~IEEE}, Sangho Lee, Hwejae Lee, and Huy Kang Kim,~\IEEEmembership{Member,~IEEE}
\thanks{Manuscript received Jan. 10, 2023; revised Jun. 7, 2023, Aug. 30, 2023, and October 18, 2023; accepted October 21, 2023.
This research was supported by the 2021 autonomous driving development innovation project of the Ministry of Science and ICT, ``Development of technology for security and ultra-high-speed integrity of the next-generation internal network of autonomous vehicles'' (No. 2021-0-01348).
\textit{(Corresponding author: Huy Kang Kim.)}}
\thanks{Seonghoon Jeong is with the Institute of Cybersecurity and Privacy, Korea University, Seoul 02841, Republic of Korea (e-mail: seonghoon@korea.ac.kr).}
\thanks{Sangho Lee is with the Samsung Research, Seoul 06765, Republic of Korea (e-mail: s35.lee@samsung.com).}
\thanks{Hwejae Lee and Huy Kang Kim are with the School of Cybersecurity, Korea University, Seoul 02841, Republic of Korea (e-mail: \{hwejae94, cenda\}@korea.ac.kr.)}
\thanks{This is the Accepted version of an article for publication in IEEE TVT. \copyright 2023 IEEE. Personal use of this material is permitted. Permission from IEEE must be obtained for all other uses, in any current or future media, including reprinting/republishing this material for advertising or promotional purposes, creating new collective works, for resale or redistribution to servers or lists, or reuse of any copyrighted component of this work in other works. Digital Object Identifier 10.1109/TVT.2023.3327275}%
}
\markboth{IEEE Transactions on Vehicular Technology,~Vol.~73,~No.~3,~2024}{Jeong \MakeLowercase{\textit{et al.}}: X-CANIDS: Signal-Aware Explainable Intrusion Detection System for Controller Area Network-Based In-Vehicle Network}

\maketitle

\begin{abstract}
Controller Area Network (CAN) is an essential networking protocol that connects multiple electronic control units (ECUs) in a vehicle. However, CAN-based in-vehicle networks (IVNs) face security risks owing to the CAN mechanisms. An adversary can sabotage a vehicle by leveraging the security risks if they can access the CAN bus. Thus, recent actions and cybersecurity regulations (\textit{e.g.,}~UNR 155) require carmakers to implement intrusion detection systems (IDSs) in their vehicles. The IDS should detect cyberattacks and provide additional information to analyze conducted attacks. Although many IDSs have been proposed, considerations regarding their feasibility and explainability remain lacking. This study proposes X-CANIDS, which is a novel IDS for CAN-based IVNs. X-CANIDS dissects the payloads in CAN messages into human-understandable signals using a CAN database. The signals improve the intrusion detection performance compared with the use of bit representations of raw payloads. These signals also enable an understanding of which signal or ECU is under attack. X-CANIDS can detect zero-day attacks because it does not require any labeled dataset in the training phase. We confirmed the feasibility of the proposed method through a benchmark test on an automotive-grade embedded device with a GPU. The results of this work will be valuable to carmakers and researchers considering the installation of in-vehicle IDSs for their vehicles.
\end{abstract}

\section{Introduction}
In-vehicle networks (IVNs) are essential for vehicles that are operated by several electronic control units (ECUs). Among the various networking protocols that have been designed for vehicles, Controller Area Network (CAN), which replaces the mesh-like wiring harness with a bus topology, is the most successful. CAN 2.0A, which was released in 1991, is currently employed in almost all vehicles because it meets the crucial requirements of IVNs. In particular, the bus topology, arbitration mechanism, and short frame enable broadcasting, interconnect multiple ECUs, and prevent medium occupation, respectively. However, these mechanisms are the root cause of security risks in CAN-based IVNs, which allow an adversary to eavesdrop on in-vehicle communication, inject arbitrary messages, and cause denial of service of a specific ECU~\cite{Cho2016a} or an entire CAN bus~\cite{LeeJK17}. 

Until the early 2010s, adversaries were considered negligible because they must have physical access to a CAN-based IVN to leverage the security risks. However, this situation has changed with the widespread use of connected vehicles, as the connectivity broadens the remote attack surfaces of connected vehicles~\cite{MillerV15, KimKJPK21}. Compromised in-vehicle infotainment systems may allow the adversaries to obtain access to IVNs remotely. Previous studies supported this concern~\cite{JoCNWL17, Gayou18, TeamFluoroacetate19, CostantinoM19, TencentMBUX21}. In particular, Miller and Valasek~\cite{MillerV15} jumpstarted cybersecurity studies on vehicles with the proof of concepts of remote hacking into a CAN-based IVN of a Jeep Cherokee.

It is crucial to protect vehicles from cyberattacks to safeguard passengers and pedestrians against unexpected vehicle behavior. Nevertheless, it is impossible to remedy the security risks of CAN-based IVNs without revising the protocol specifications. Intrusion detection systems (IDSs) have been proposed to identify anomalies in CAN-based IVNs~\cite{KimKJPK21, JoC21}. In the early research stages, statistical approaches and conventional machine-learning algorithms were considered. As deep-learning techniques have evolved, recent studies have tended to adapt such techniques for precise intrusion detection. IDSs are currently becoming an essential component of vehicles. For example, United Nations Regulation No. 155 (UNR 155), which is a recent automotive cybersecurity regulation that will take effect in many countries from 2024, states that \textit{``the vehicle manufacturer shall implement measures for the vehicle type to (a) detect and prevent cyber-attacks against vehicles of the vehicle type; (b) support the monitoring capability of the vehicle manufacturer with regards to detecting threats; (c) provide data forensic capability to enable an analysis of attempted or successful cyber-attacks''} (see \S 7.3.5 in \cite{UNR155}).

In this study, we propose a practical IDS named X-CANIDS to address the following three limitations with respect to IDSs for CAN-based IVNs. 
First, previous IDSs did not provide additional information for forensics. Most carmakers have their own pattern database to distinguish a malfunction. 
An explanation of detection result can help carmakers analyze conducted attacks and prepare a remedy such as updating their database. Supervised methods (\textit{e.g.,}~\cite{HossainIOFK20, TariqLKW20}) have been designed to distinguish the type of attack. However, they require ground-truth labels and only distinguish predefined attack types.
Second, the evaluations have lacked a feasibility perspective. Evaluation is necessary because an IDS works on an ECU or an in-vehicle component. Regardless of the effectiveness of an IDS, it may be useless owing to bottlenecks.
Finally, limited studies have considered the use of signals rather than raw payloads. Signals help to improve the detection performance of an IDS because they reflect the context of the vehicle. Nevertheless, the use of signals has rarely been discussed because of a lack of knowledge regarding payload deserialization methods. To date, two studies have used sensor values through the on-board diagnostics (OBD)-II feature~\cite{WasicekPWBS17} and several reverse-engineered signals~\cite{ShahriarXMLH22}.

The contributions of this study are summarized as follows:

\BfPara{1. Self-supervised intrusion detection with signals} We propose X-CANIDS, which is a novel method that consists of a feature generator and intrusion detection model. The feature generator builds a time-series representation of signals that are deserialized from the payloads of the CAN messages. We use a CAN database to train X-CANIDS with 107 signals. The detection model is trained using an attack-free dataset. X-CANIDS can detect zero-day attacks, of which we are unaware at the time of implementation.

\BfPara{2. Explainability} X-CANIDS provides additional information on which systems (\textit{i.e.,}~ECUs) and what data were compromised, even if no ground-truth labels or intrusion datasets are available in the training phase. The explainability is beneficial for carmakers and incident response teams to analyze conducted attacks. To this end, we leverage the signalwise reconstruction error combined with an autoencoder.

\BfPara{3. Feasibility} Our method is benchmarked on our NVIDIA Jetson AGX Xavier, which is an automotive-grade embedded device. The benchmark results confirm that X-CANIDS is promising for real-world use cases. X-CANIDS achieves a deterministic detection latency of 38.2512--73.2512 ms on the embedded device with a feature generation frequency and batch size of 200 Hz and 8, respectively.

The remainder of this paper is organized as follows. 
In \autoref{sec:preliminaries}, we provide the background for this study.
In \autoref{sec:methodology}, we describe the proposed X-CANIDS.
\autoref{sec:dataset} outlines the CAN datasets that are used in the experiment.
The experimental results are presented and the detection performance, explainability, and feasibility are discussed in \autoref{sec:experimental_result}. 
In \autoref{sec:related_work}, we categorize prior studies based on the features that are used to detect in-vehicle intrusion.
Finally, we conclude the study in \autoref{sec:conclusion}.

\section{Preliminaries}
\label{sec:preliminaries}
\subsection{Terminology}
We present the terminology used in this paper.
The \textit{payload} is a bit sequence that is encapsulated in the data field of a CAN frame (see \autoref{fig:CAN_frame_structure}). The \textit{signal} indicates a sensor value that is deserialized from a portion of the payload. A \textit{stream} is a set of CAN messages with a particular arbitration ID. The aim of this study is to develop an \textit{explainable} and a \textit{feasible} IDS. We can state that the proposed IDS is explainable when a detection result allows us to understand which signal or ECU is affected by attacks. We can state that the proposed IDS is feasible when it can evaluate all given inputs within a deterministic latency on an ECU or an automotive-grade embedded device.

\subsection{CAN Frame}
\begin{figure}[t]
\centering
\includegraphics[width=\linewidth]{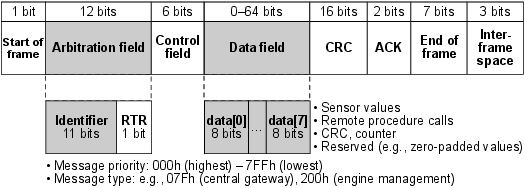}
\caption{CAN 2.0A frame structure. An ECU application refers to the arbitration and data field.}
\label{fig:CAN_frame_structure}
\end{figure}

The CAN frame is a communication unit between two in-vehicle ECUs that are connected by a CAN bus. \autoref{fig:CAN_frame_structure} depicts the CAN 2.0A frame structure. Among the fields, the ECU firmware uses the 11-bit arbitration identifier (AID) field and a 64-bit data field. The AID is considered as a categorical value for determining the message type as well as a numerical value that represents the priority of the message. The data field contains a payload that represents various signals that are used to operate vehicles, such as sensor values, Booleans, inter-ECU procedure calls, cyclic redundancy check values, sequential counters, and zero padding. As a CAN frame does not contain a transmission timestamp, the CAN receiver assigns a timestamp for each inbound CAN message using its internal clock. In the ECU firmware, a transmitted CAN frame is represented as a CAN message $m \rightarrow (t, a, {\bf p})$, where the timestamp $t \geq 0$, AID $a\in\{0, 1, \cdots, 2047\}$, and bit sequence vector of the payload ${\bf p}=\{p_i\ | p_i \in \{0, 1\}~\text{for}~i=1..n\}$, with $n \in \{0, 8, 16, \cdots, 64\}$.

The ECU firmware serializes one or more signals in the data field prior to transmission. Each receiver can deserialize the payload of the data field to use the original values. Researchers can easily obtain a CAN dataset $\mathcal{M}=\{m_1, m_2, m_3, \cdots\}$ through their CAN nodes by leveraging the broadcast nature of the CAN. However, it is difficult for them to understand the specific meaning of $m$ owing to the lack of information regarding the exact representation of the given $a$ and ${\bf p}$. Meanwhile, researchers who wish to build after-market autonomous driving kits or use in-vehicle signals for specific purposes, such as building a payload-based IDS for CAN buses~\cite{ShahriarXMLH22}, are motivated to reverse engineer CAN message payloads. Because manual reverse engineering requires substantial effort, several automated methods have been proposed for this purpose~\cite{VermaBSHI21, YoungSZ20, PeseSCNCS19, KangSJK18, VermaBH18, HuybrechtsVBBH17, MarkovitzW17}, and  increasingly sophisticated results have been achieved in recent years. However, these methods remain insufficient for deserializing hundreds of signals precisely.

\subsection{CAN Database}
\label{subsec:can_database}
The CAN database is a network dissector that consists of formal payload deserialization descriptions. We introduce the CAN database using a straightforward example, because signals that are deserialized from CAN messages are crucial inputs for the proposed method. A CAN database describes the specification of a certain CAN-based IVN, including the bitrate, list of ECUs, signals, and Tx-Rx ECU relationships. The CAN database specifies the bit indices, endianness, signness, scale, offset, value range, units, and list of ECUs that refer to each signal. In general, CAN databases are composed by carmakers at the time of the IVN design.

\begin{figure}[!t]
\centering
\includegraphics[width=\linewidth]{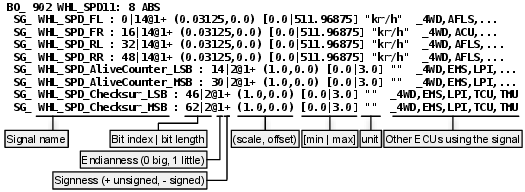}
\caption{Snippet of CAN database \textit{hyundai\_2015\_ccan.dbc}~\cite{opendbc}.}
\label{fig:snippet_CAN_database}
\end{figure}

Recently, Comma.ai, which is a company that develops after-market autonomous driving kits, released CAN databases that work with some commercialized vehicles in a public Git repository known as OpenDBC~\cite{opendbc}. The CAN databases are formatted in the well-known DBC file format that was introduced by Vector Informatik GmbH. \autoref{fig:snippet_CAN_database} presents a snippet of a CAN database that was obtained from the repository. The first line states that an ECU named ABS (\textit{i.e.,}~anti-lock braking system) transmits the message WHL\_SPD\_11, which consists of $a=902$ and $|{\bf p}|=64$ (\textit{i.e.,}~8 bytes). The remainder defines eight signals of rotation speeds (front left, front right, rear left, and rear right wheels) that are represented in km/h units, two checksum values, and two alive counters. The meanings of the CAN database syntax are annotated at the bottom of the figure. 

\newcommand\mycommfont[1]{\footnotesize\ttfamily\textcolor{blue}{#1}}
\SetCommentSty{mycommfont}
\SetKwInput{KwInput}{Input}                
\SetKwInput{KwOutput}{Output}              

\begin{algorithm}[t]
\small
\DontPrintSemicolon

\KwInput{$a, {\bf p}$: AID and payload of CAN message}
\KwData{${\rm DBC}$: CAN database} 
\KwOutput{${\bf s}=\{s_i |{i=1..n}\}$: List of deserialized signals}

signals $\leftarrow$ get\_signal\_specification(DBC, $a$) \;
$n \leftarrow |{\rm signals}|$ \tcp*{Number of signals}
Initialize a new vector ${\bf s} \in \mathbb{R}^{|n|}$ \;
\For{$i\leftarrow1$ \KwTo $n$}
{
    \tcp{Parse the specification of a signal}    
bit\_idx, bit\_len, endianness, signness, scale, offset, min, max $\leftarrow {\rm signal}_{i}$  \;
    \tcp{Obtain a subset of bit sequence}
    ${\bf p}' \leftarrow \{p_{\rm bit\_idx}, \cdots, p_{\rm bit\_idx+bit\_len} \}$ \;    
    \tcp{Decode the bit sequence into an integer}
    $s_i \leftarrow {\rm int}({\bf p}', {\rm endianness}, {\rm signness})$ \;
    \tcp{Scale the signal}
    $s_i \leftarrow s_i / {\rm scale} + {\rm offset}$ \;
    \If{not ${\rm min} \leq s_i \leq {\rm max}$}
    {
        Raise an error. \;
    }
}
\caption{Deserialization of a CAN message.}
\label{algo:deserialization}
\end{algorithm}

The deserialization procedure $D$ is described in \autoref{algo:deserialization}. The deserialization procedure can be represented by $D(a, {\bf p}) = {\bf s} = \{s_i | s_i \in \mathbb{R}~\text{for}~i=1..n\}$, where $n$ is the number of signals that are defined in the given DBC (see lines 1--2). The output vector {\bf s} contains human-understandable signals. The loop (lines 4--10) can be executed concurrently because $s_i$ are independent of one another. 

\subsection{Adversary and Attack Model}
\label{subsec:adversary}
\begin{figure}[!t]
  \centering
  \includegraphics[width=\linewidth]{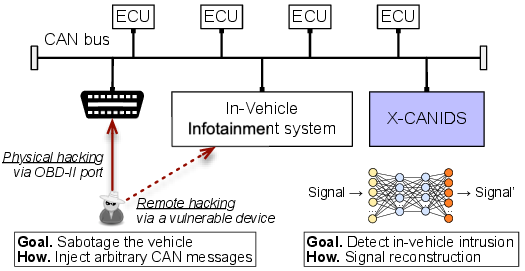}
  \caption{Considered in-vehicle network architecture.} 
  \label{fig:diagram}
\end{figure}

In this study, we consider an adversary who wants to sabotage a vehicle by injecting arbitrary CAN messages. \autoref{fig:diagram} depicts the supposed adversary in the CAN-based IVN. The adversary must obtain access to the target CAN-based IVN to conduct an attack. The adversary may consider physical hacking by installing a CAN dongle at the OBD-II port. The attacker may also consider the remote exploitation of vehicle-to-everything communication-enabled ECUs, such as an infotainment system~\cite{Gayou18, TeamFluoroacetate19, CostantinoM19, TencentMBUX21}. Once the adversary obtains access, they can conduct five types of attacks ~\cite{Cho2016a, ShahriarXMLH22}, as follows:

\subsubsection{Fuzzing Attack} It manipulates various ECUs with random payloads and it can be performed with CAN messages that contain random AIDs and payloads. The attack can cause a malfunction of the target vehicle even if the adversary does not have prior knowledge of the in-vehicle communications.
\subsubsection{Fabrication Attack} A specific ECU is manipulated as the intention of the adversary, and it can be performed using well-crafted CAN messages with a specific AID and payload. As a legitimate ECU periodically transmits CAN messages with the same AID, an adversary can transmit their CAN message directly after every benign message.
\subsubsection{Suspension Attack} It neutralizes an ECU by exploiting the error-handling mechanism of the CAN~\cite{Cho2016a}. A target ECU does not transmit any CAN messages during the attack.
\subsubsection{Masquerade Attack} It is a combination of the fabrication and suspension attacks. A stream from a specific ECU is replaced with arbitrary messages that are generated by the adversary during the attack.
\subsubsection{Replay Attack} An adversary captures legitimate CAN messages in a certain period. Then, they transmit the CAN messages within the CAN bus. The attack can cause a certain malfunction that the target vehicle have performed in the capture duration.

\section{Methodology}
\label{sec:methodology}

\begin{figure*}[!t]
  \centering
  \includegraphics[width=\linewidth]{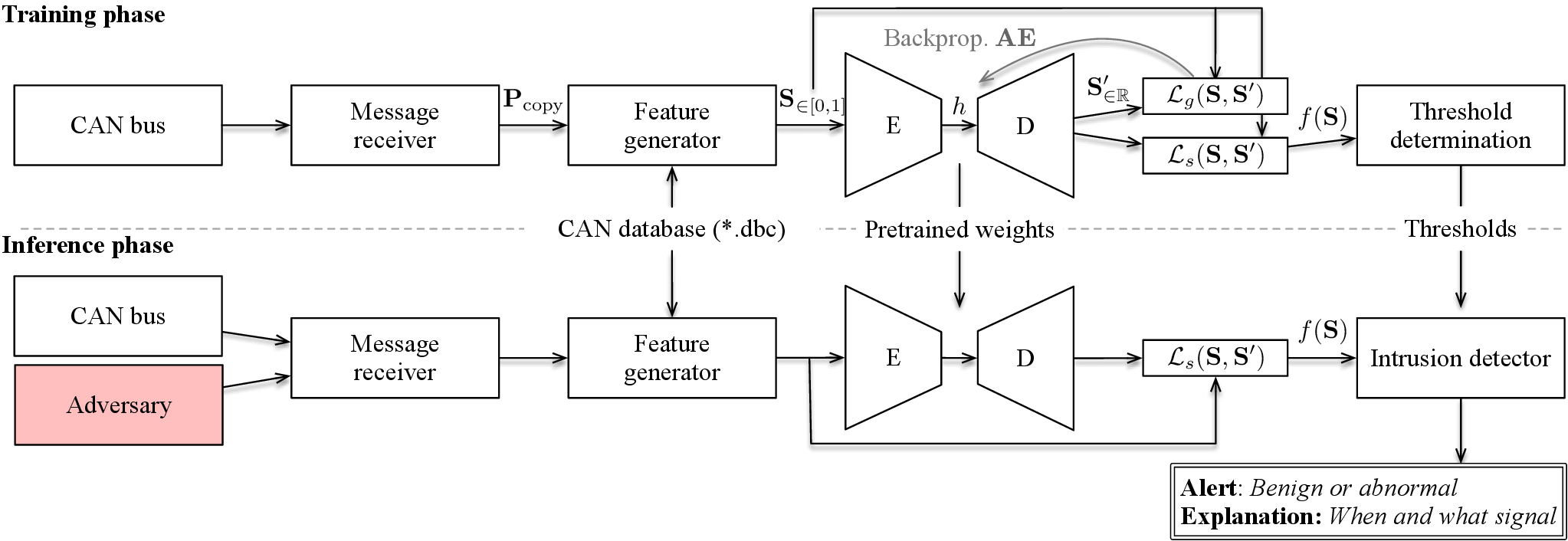}
  \caption{Proposed framework. In the training phase, the proposed framework uses attack-free CAN messages to train the autoencoder and to determine the threshold. In the inference phase, the proposed framework determines the signal that is affected by the adversary if the CAN bus is under attack.}
  \label{fig:framework}
\end{figure*}

In this section, we present X-CANIDS, which consists of a message receiver, a feature generator, an autoencoder, and a decision-maker. As illustrated in \autoref{fig:diagram}, the proposed in-vehicle IDS is directly connected to the CAN-based IVN to receive all CAN messages instantly. It is a preferable architecture for prior CAN IDSs, while maintaining a simple topological structure. \autoref{fig:framework} outlines the proposed framework. In the training phase, X-CANIDS uses benign CAN messages to train the autoencoder ${\bf AE}$ towards small global errors. At the end of the training phase, X-CANIDS determines the thresholds using signalwise errors. In the inference phase, X-CANIDS determines whether a given input is affected by an adversary using pretrained weights and thresholds. Signalwise errors are considered to explain the detected attack. The CAN database is required in both phases for the feature generator to deserialize the signals from the CAN messages.

\subsection{Message Receiver}
The message receiver is connected to the CAN bus to monitor all CAN messages. The message receiver contains the matrix ${\bf P}\in \{\emptyset, 0, 1\}^{|N \times M|}$ to cache the latest payload of each stream, where $N$ is the number of streams in the CAN bus and $M$ is the maximum length of the payload (64 bits for the CAN 2.0A bus and 512 bits for the CAN-FD bus).

\begin{equation} \label{eq:P}
  {\bf P} = 
    \begin{pmatrix}
    {\bf p}_1 \\
    {\bf p}_2 \\
    \vdots \\
    {\bf p}_{N}
    \end{pmatrix} 
    =
    \begin{pmatrix}
    p_{11} & p_{12} & \cdots & p_{1M} \\
    p_{21} & p_{22} & \cdots & p_{2M} \\
    \vdots  & \vdots  & \ddots & \vdots  \\
    p_{N1} & p_{N2} & \cdots & p_{NM} \\ 
    \end{pmatrix}
\end{equation}
In \cref{eq:P}, the $n$-th row of the matrix ${\bf P}$ represents the latest payload of a certain stream for each ${\bf p}_{n}$. Initially, $p_{n, m}=\emptyset \forall_{n\in\{1..N\},m\in\{1..M\}}$. As the message receiver monitors each CAN message from the CAN bus, $p_{n, m}$ becomes 0 or 1. It should be noted that ${\bf P}$ is volatile because the elements change continuously upon the arrival of CAN messages.

\subsection{Feature Generator}
\label{subsec:feature_generator}
The feature generator interprets each ${\bf P}$ into a feature matrix ${\bf S}$, which is fed to the autoencoder. The feature generator pipeline consists of the following: (1) the payload sampler, (2) deserializer, (3) feature scaler, and (4) time-series feature generator. 

\subsubsection{Payload Sampler}
\label{subsec:payload_sampler}
The payload sampler captures ${\bf P}_{\rm copy}$, which is a static copy of ${\bf P}$, from the message receiver in every time interval $t$. The payload sampler begins working when ${\bf P}$ satisfies $p_{n,1}\neq\emptyset \forall_{n\in\{1..N\}}$, which means that the message receiver observes each stream at least once. Each copy is forwarded to the deserializer.

\subsubsection{Deserializer}
The deserializer converts a given ${\bf P}_{\rm copy}$ into a vector ${\bf s} \in \mathbb{R}$ that contains human-understandable signals, such as the engine speed and steering wheel angle. The deserialization procedure for ${\bf P}_{\rm copy}$ can be represented as

\begin{equation} \label{eq:deserialization_payload}
  {\bf P}_{\rm copy} = 
    \begin{pmatrix}
      {\bf p}_1 \\
      {\bf p}_2 \\
      \vdots \\
      {\bf p}_{N} \\
    \end{pmatrix}
    \begin{matrix}
      \xrightarrow{D(a_1, {\bf p}_1)} \\
      \xrightarrow{D(a_2, {\bf p}_2)} \\
      \xrightarrow{\text{deserialize}} \\
      \xrightarrow{D(a_N, {\bf p}_N)} \\
    \end{matrix}
    \begin{pmatrix}
      {\bf s}_1 \\
      {\bf s}_2 \\
      \vdots \\
      {\bf s}_{N} \\
    \end{pmatrix}
    \xrightarrow{\text{concatenate}}
    {\bf s}.
\end{equation}

The streamwise deserialization procedure calculates \autoref{algo:deserialization}. Thus, the output vectors ${\bf s}_1, {\bf s}_2, \cdots, {\bf s}_n$ are dependent on the given CAN database and ${\bf P}_{\rm copy}$. In particular, the CAN database determines the number of elements in ${\bf s}_i$. The vector ${\bf s}=\Vert_{n=1}^{N}{\bf s}_n$ represents the concatenation of all deserialized signals and is fed to the feature scaler.

\subsubsection{Feature Scaler} 
The feature scaler is crucial because the elements of ${\bf s}$ have various value ranges. For example, the engine speed and steering wheel angle can be represented within $[0, 8191]$ RPM and $[-1024, 1023]$ degrees, respectively. Furthermore, a binary value (0 or 1) can be used to represent whether a foot brake is engaged. The variation in signal ranges may result in unstable training of $\bf{AE}$ because it can be considered as a weight (\textit{i.e.,}~feature importance).

We design a lightweight feature scaler to achieve robust performance. The signal scaler is designed to normalize a given ${\bf s}$ to ${\bf\hat s}=\{{\hat s_i}|{\hat s_i}\in[0, 1], {i=1..x}\}$, where $x=|{\bf s}|$ is the number of concatenated signals. The proposed feature scaling procedure is presented in \cref{eq:signal_scaling}, where the parameters ${\rm min}_i$ and ${\rm max}_i$ are the minimum and maximum values, respectively, of the $i$-th signal described in the CAN database.

\begin{equation}
  \label{eq:signal_scaling}
	{\hat s_i} = \frac{s_i-{\rm min}_i}{{\rm max}_i-{\rm min}_i}
\end{equation}

Our feature scaler is the same as the conventional min-max scaler, except for the determination method of the parameters ${\rm min}_i$ and ${\rm max}_i$. If we had applied the min-max scaler, it would fit the parameters according to observations in the training phase (\textit{i.e.,}~a training set). A downside of the min-max scaler is that it does not correctly handle outliers that are over the maximum or under the minimum values that may be observed in the inference phase. We revised the min-max scaler to this end. Our feature scaler leverages the minimum and maximum values that are predefined in the CAN database to overcome this drawback. Note that a scaled signal ${\hat s_i}$ from our feature scaler is always represented by 0--1 because $s_i$ satisfies ${\rm min}_i \leq s_i \leq {\rm max}_i$ (cf. lines 9--10 of \autoref{algo:deserialization}). The scaled vector ${\bf\hat s}$ is fed to the time-series feature generator.

\subsubsection{Time-Series Feature Generator}
\label{subsec:time_series_feature_generator}
The time-series feature generator builds an input for the autoencoder in a sliding-window manner. It temporarily remembers the most recent $w$ input vectors and returns a two-dimensional matrix ${\bf S} \in [0, 1]^{w \times x}$ by stacking them. The window size $w$ is a parameter that determines the number of time steps that are contained by the feature. The time-series feature generator returns ${\bf S}$ every $t$ as the payload sampler feeds ${\bf P}_{\rm copy}$. The matrix representation helps the autoencoder to understand the time series and lateral relationships between the signals.

\subsection{Autoencoder}
In the proposed framework, the autoencoder ${\bf AE}$ is adapted to model the attack-free state of a moving vehicle. The autoencoder is a self-supervised neural network that consists of an encoder and a decoder, as expressed by 

\begin{equation}\label{eq:autoencoder}
  {\bf AE}({\bf S}) = {\rm Decoder}({\rm Encoder}({\bf S}))= {\rm Decoder}(h)={\bf S}'
\end{equation}
The encoder compresses the input feature ${\bf S}$ into a low-dimensional latent vector $h$. Subsequently, the decoder attempts to reconstruct the original input data as far as possible using the latent vector. In the training phase, ${\bf AE}$ is fitted with features from benign datasets using backpropagation to reduce the global mean squared error (MSE). The global MSE is calculated as follows:
\begin{equation}\label{eq:mse_global}
  \mathcal{L}_g({\bf S}, {\bf S}') = \frac{1}{wx} \sum_{j=1}^w\sum_{i=1}^x ({\bf S}_{ji}-{\bf S}_{ji}')^2.
\end{equation}
It is also referred to as the reconstruction error in terms of an autoencoder. The goal of ${\bf AE}$ is to exhibit a small reconstruction error in the inference phase, particularly when a given sample is attack free. However, ${\bf AE}$ is required to exhibit a high reconstruction error when the input ${\bf S}$ is affected by an adversary.

${\bf AE}$ is supposed to be computed on an automotive-grade embedded device. Therefore, the model complexity should be considered while minimizing the reconstruction error. We conceive the six candidate layers for ${\bf AE}$ as follows: the fully connected, 1D and 2D convolutional, 1D separable convolutional, long short-term memory (LSTM), and bidirectional LSTM (BiLSTM) layers. We evaluate the layers using training and validation datasets.

\subsection{Intrusion Detection and Explanation}
The global MSE can be used solely to distinguish anomalies. Nevertheless, we measure the signalwise MSE to obtain explainable intrusion detection results. In this section, we define the inference function $f({\bf S})$ that calculates the signalwise MSE. Thereafter, we introduce the threshold determinator and intrusion detector.

\subsubsection{Inference Function}
Once ${\bf AE}$ has been fitted, the signalwise loss function $\mathcal{L}_s$ is combined with ${\bf AE}$ to formulate the inference function $f({\bf S})$. Equation \cref{eq:detection_function} presents the inference function that returns the loss vector ${\bf l}$. The element $l_i$ is the loss of the $i$-th signal.

\begin{equation}\label{eq:detection_function}
  \begin{aligned}
  f({\bf S}) = & \mathcal{L}_s({\bf S}, {\bf AE}({\bf S})) = \mathcal{L}_s({\bf S}, {\bf S}') \\ = 
 & \frac{1}{w} \sum_{j=1}^w({\bf S}_j-{\bf S}_j')^2 = {\bf l} = \{l_1, l_2, \cdots, l_x\}
  \end{aligned}
\end{equation}

\subsubsection{Threshold Determination}
The threshold determinator is used during the training phase. This module aims to determine $\theta_i$ for the error rate calculation of the $i$-th signal and the detection threshold $\Theta$ to raise the alarm. 

First, the module calculates a set of loss vectors $\{\bf{l}_1, \bf{l}_2, \cdots\}$ using the entire training set. Second, $\theta_i = \overline{l_i} + 3\sigma_i$ is considered for the $i$-th signal, where $\overline{l_i}$ and $\sigma_i$ are the mean and standard deviation of $l_i$s in the set, respectively. Third, the module measures a set of error-rate vectors $\{\bf{r}_1, \bf{r}_2, \cdots\}$ using the entire validation set. The error rate vector {\bf r} is derived as follows:
\begin{equation}\label{eq:error_rate}
  {\bf r} = \{r_i | r_i = l_i/\theta_i~\text{for}~ i=1..x\}.
\end{equation}
Finally, $\Theta$ is determined by the $q$-th percentile of ${\rm max}({\bf r})$ for all ${\bf r}$s in the set, where $0.95 \leq q \leq 1$ and $q$ is a hyperparameter that determines the detection sensitivity.

\subsubsection{Intrusion Detection and Explanation}
The intrusion detection module uses the $\theta_i$s and $\Theta$ from the threshold determinator at the beginning of the inference phase. The intrusion detector obtains ${\bf r}$ for each ${\bf S}$ using \cref{eq:detection_function} and \cref{eq:error_rate}. The module raises an alarm if ${\bf r}$ satisfies $\rm{max}({\bf r}) > \Theta$. If an intrusion is identified, the intrusion detector identifies the affected signal index $i$ using $\rm{argmax}({\bf r})$.

\section{Datasets}
\label{sec:dataset}
\begin{table}[t]
  \caption{List of CAN datasets.} 
  \label{table:list_dataset}
  \centering
  \begin{tabular}{rrrll}
  \hline
  Dataset & \# CAN msg. & Duration & Context & Used for\\
  \hline
  $\mathcal{M}_1$ &  3,123,784 &  23 m 52 s &   Driving (18.7 km) & Training \\
  $\mathcal{M}_2$ &  4,134,495 &  31 m 35 s &   Driving (19.3 km) & Training \\
  $\mathcal{M}_3$ &  3,233,752 &  24 m 42 s &   Driving (19.4 km) & Training \\
  $\mathcal{M}_4$ &  4,761,315 &  36 m 23 s &   Driving (21.7 km) & Training \\\
  $\mathcal{M}_5$ &  2,915,969 &  22 m 17 s &   Driving (18.8 km) & Validation \\
  $\mathcal{M}_6$ &  4,279,902 &  32 m 42 s &   Driving (18.8 km) & Testing  \\
  $\mathcal{M}_7$ &  4,817,589 &  36 m 54 s &   Stationary (0 km) & Training \\\hline
  \end{tabular}
\end{table}

The datasets used in the experiment are discussed in this section. Publicly available CAN intrusion datasets~(\textit{e.g.,}~\cite{LeeJK17, VermaIBHMKC22}) exist. However, they could not be utilized because the datasets were stationary or prepared using a dynamometer. Therefore, we captured CAN messages from the Hyundai LF Sonata 2017. We used a Kvaser Memorator Pro 2xHS to leverage the high-precision embedded clock while capturing the in-vehicle CAN messages. The device was connected to our vehicle through an OBD-II port, which allowed us to access the chassis--CAN bus. We prepared the seven CAN datasets that are listed in \autoref{table:list_dataset}. We captured six datasets while driving a vehicle on urban roads in Seoul, Republic of Korea. One dataset was captured while the vehicle was stationary. The datasets are listed in chronological order. For instance, $\mathcal{M}_2$ was collected after we have collected $\mathcal{M}_1$. 

As shown in \autoref{fig:framework}, X-CANIDS consists of the training and inference phases. In the training phase, pre-captured datasets are used to train an autoencoder. In the inference phase, the trained model processes a live stream. In this paper, we try not to merge entire datasets and conduct an N-fold cross-validation, in which subsets of $\mathcal{M}_i \forall_{i \in \{1..6\}}$ are used in both training and testing. Instead, we selected old datasets $\mathcal{M}_1$--$\mathcal{M}_4$ as the training set, new dataset $\mathcal{M}_5$ as the validation set, and the newest $\mathcal{M}_6$ as the test set. We believe such a selection can reflect a real-world use case.

We used the CAN database \textit{hyundai\_2015\_ccan.dbc} that is available on OpenDBC~\cite{opendbc} to deserialize the payloads of the datasets into signals. We manually confirmed that the CAN database allowed us to acquire appropriate sensor values, such as the steering wheel angle, temperature, velocity, tire pressure, and door open state.

\subsubsection{Overview of Datasets}
\begin{table}[t]
  \caption{Summary of $\mathcal{M}_1$ that consists of 62 streams.}
  \label{table:dataset_summary}
  \centering
  \begin{adjustbox}{width=\linewidth,center}
  \setlength{\tabcolsep}{4pt} 
  \begin{tabular}{rlrrrrr}
  \hline
  AID &           Sender ECU  &  Mean $\Delta t$ &  Std. $\Delta t$ &  DLC &  \# signals &  \# uniq. ${\bf p}$ \\\hline
  042h &        DATC12 &          1.00 &      0.000261 &    8 &         7 &              1 \\
  043h &        DATC13 &          1.00 &      0.000261 &    8 &        24 &              1 \\
  044h &        DATC11 &          0.96 &      0.180892 &    8 &         6 &             14 \\
  07Fh &          CGW5 &          1.00 &      0.000124 &    8 &        25 &              1 \\
  080h &     EMS\_DCT11 &          0.01 &      0.000304 &    8 &        10 &        105,446 \\\hline
  081h &     EMS\_DCT12 &          0.01 &      0.000356 &    8 &         6 &          1,000 \\
  111h &         TCU11 &          0.01 &      0.000191 &    8 &        13 &            555 \\
  112h &         TCU12 &          0.01 &      0.000201 &    8 &        12 &          6,083 \\
  113h &         TCU13 &          0.01 &      0.000203 &    8 &        18 &            441 \\
  153h &         TCS11 &          0.01 &      0.000132 &    8 &        29 &             15 \\\hline
  162h &     TCU\_DCT13 &          0.01 &      0.000207 &    3 &         3 &          6,185 \\
  164h &         VSM11 &          0.01 &      0.000220 &    4 &         6 &             16 \\
  18Fh &       EMS\_H12 &          0.01 &      0.000449 &    8 &        21 &          2,117 \\
  200h &         EMS20 &          0.01 &      0.000370 &    6 &         3 &            348 \\
  220h &         ESP12 &          0.01 &      0.000211 &    8 &        14 &        130,706 \\\hline
  251h &        MDPS12 &          0.01 &      0.000311 &    8 &        11 &        116,572 \\
  260h &         EMS16 &          0.01 &      0.000664 &    8 &        15 &         41,585 \\
  2B0h &         SAS11 &          0.01 &      0.000683 &    5 &         5 &         17,265 \\
  316h &         EMS11 &          0.01 &      0.000823 &    8 &        13 &         63,371 \\
  329h &         EMS12 &          0.01 &      0.000352 &    8 &        19 &          7,776 \\\hline
  381h &        MDPS11 &          0.02 &      0.000530 &    8 &        13 &         32,040 \\
  383h &        FATC11 &          0.02 &      0.000395 &    8 &        19 &            320 \\
  386h &     WHL\_SPD11 &          0.02 &      0.000399 &    8 &         8 &         61,540 \\
  387h &     WHL\_PUL11 &          0.02 &      0.000418 &    6 &         9 &         60,269 \\
  410h &      CGW\_USM1 &          0.20 &      0.000522 &    8 &        17 &              1 \\\hline
  436h &         PAS11 &          0.05 &      0.000661 &    4 &        12 &              1 \\
  47Fh &         ESP11 &          0.02 &      0.000256 &    6 &        11 &            256 \\
  490h &         EPB11 &          0.05 &      0.000515 &    7 &        14 &              1 \\
  492h &         EMS19 &          0.05 &      0.000460 &    8 &        13 &              4 \\
  4F1h &         CLU11 &          0.02 &      0.000446 &    4 &        12 &         14,476 \\\hline
  500h &         ACU14 &          0.10 &      0.000504 &    1 &         3 &              1 \\
  502h &         TCU14 &          0.10 &      0.000374 &    4 &         7 &              1 \\
  507h &         TCS15 &          0.10 &      0.000288 &    4 &        11 &              1 \\
  50Ch &         CLU13 &          0.10 &      0.000463 &    8 &        17 &          1,484 \\
  520h &          CGW3 &          0.10 &      0.000850 &    8 &         4 &              1 \\\hline
  522h &   GW\_IPM\_PE\_1 &          0.20 &      0.000558 &    8 &        10 &              1 \\
  52Ah &         CLU15 &          0.10 &      0.076209 &    8 &        15 &             94 \\
  533h &           --- &          0.10 &      0.000706 &    8 &      --- &             29 \\
  534h &           --- &          0.10 &      0.000704 &    8 &      --- &            517 \\
  535h &           --- &          0.10 &      0.000674 &    8 &      --- &          1,249 \\\hline
  541h &          CGW1 &          0.10 &      0.014478 &    8 &        43 &              4 \\
  544h &           --- &          0.20 &      0.000422 &    8 &      --- &              1 \\
  545h &         EMS14 &          0.10 &      0.000335 &    8 &         8 &            234 \\
  547h &         EMS15 &          0.10 &      0.000545 &    8 &        12 &             38 \\
  549h &         BAT11 &          0.10 &      0.000570 &    8 &         9 &          6,635 \\\hline
  54Ch &     TCU\_DCT14 &          0.20 &      0.000385 &    8 &         2 &              1 \\
  553h &          CGW2 &          0.20 &      0.013671 &    8 &        41 &              6 \\
  555h &        FPCM11 &          0.10 &      0.000507 &    8 &         9 &            670 \\
  556h &   EngFrzFrm11 &          0.10 &      0.000578 &    8 &         6 &         11,649 \\
  557h &   EngFrzFrm12 &          0.10 &      0.000587 &    8 &         6 &          4,932 \\\hline
  559h &          CGW4 &          0.20 &      0.000655 &    8 &        23 &              1 \\
  57Fh &  HU\_MON\_PE\_01 &          2.00 &      0.000782 &    8 &         1 &              1 \\
  587h &         TMU11 &          0.20 &      0.000347 &    8 &         8 &             15 \\
  58Bh &         LCA11 &          0.10 &      0.001104 &    8 &        18 &              7 \\
  593h &        TPMS11 &          0.20 &      0.000554 &    6 &        12 &             11 \\\hline
  5A0h &         ACU11 &          1.00 &      0.001139 &    8 &        14 &              2 \\
  5B0h &         CLU12 &          1.00 &      0.000780 &    4 &         1 &            184 \\
  5B4h &           --- &          1.00 &      0.000776 &    8 &      --- &              1 \\
  5BEh &           --- &          1.00 &      0.000834 &    8 &      --- &              1 \\
  5C0h & GW\_Warning\_PE &          1.00 &      0.000810 &    8 &         7 &              1 \\\hline
  5D3h & HU\_DATC\_PE\_00 &          1.00 &      0.000806 &    8 &         3 &              1 \\
  5FAh &         ODS11 &          1.00 &      0.000965 &    8 &        10 &              1 \\\hline
  \end{tabular}
  \end{adjustbox}
  \end{table}

The dataset $\mathcal{M}_1$ is summarized in \autoref{table:dataset_summary} to introduce our CAN datasets. The chassis--CAN bus consisted of 62 unique streams. We denote the name of the transmitting ECU using the CAN database for each stream. The ECU names; for example, engine management system (EMS), transmission control unit (TCU), and motor-driven power steering system (MDPS), imply that the streams take charge of the communications among critical vehicular applications. 

We measured the mean and standard deviation of the message intervals for each stream. ECUs transmit their messages periodically through a well-known mechanism. Thus, it can be observed that the mean time intervals of the streams were approximately one of 0.01, 0.02, 0.05, 0.1, 0.2, 1, and 2 s. However, several streams consisted of nonperiodic messages, namely 044h, 52Ah, 541h, and 553h (where the standard deviation $ \geq 0.01$). These streams may not be covered by time-interval-based intrusion detection methods (\textit{e.g.,}~\cite{SongKK16, TomlinsonBSK18, OlufowobiYZB19, OlufowobiYZB19, YoungOBZ19}) that leverage the periodicity.

The number of signals that were defined in each stream is also summarized. In total, there were 688 types of signals. Furthermore, the number of unique payloads for each stream is presented. Interestingly, although we captured $\mathcal{M}_1$ while we drove the vehicle for more than 18 km, certain streams (\textit{e.g.,}~042h, 043h, and 5D3h) had a static payload.

\subsubsection{Payload Dynamics}
We expected that the payloads in the CAN messages would change more dynamically when the vehicle moved. To prove this hypothesis, we calculated the bitwise Hamming distance vector ${\bf d}$ for each stream, as follows:

\newcommand{\legendsquare}[1]{\textcolor[RGB]{#1}{\rule{1.5ex}{1.5ex}}}
\begin{figure*}[!t]
\centering
\includegraphics[width=.9\linewidth]{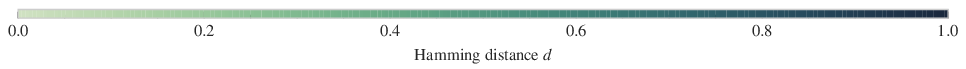}\\\vspace{-.2cm}
\subfloat[Stationary dataset $\mathcal{M}_7$ with 764 bits flipped. $\Sigma\Sigma {\bf d}=105.91$.]{\includegraphics[width=.49\linewidth]{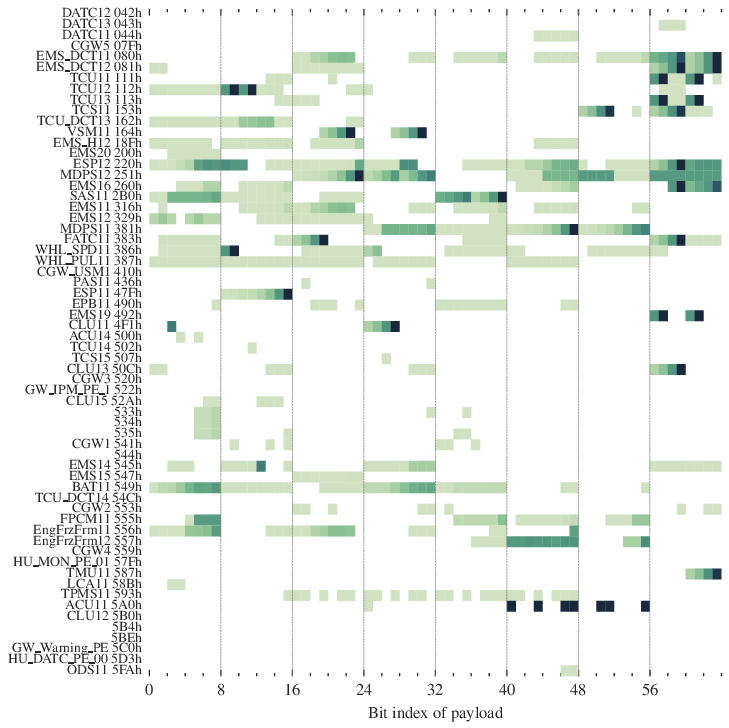}
}
\hfil
\subfloat[Driving dataset $\mathcal{M}_1$ with 968 bits flipped. $\Sigma\Sigma {\bf d}=135.45$.]{\includegraphics[width=.49\linewidth]{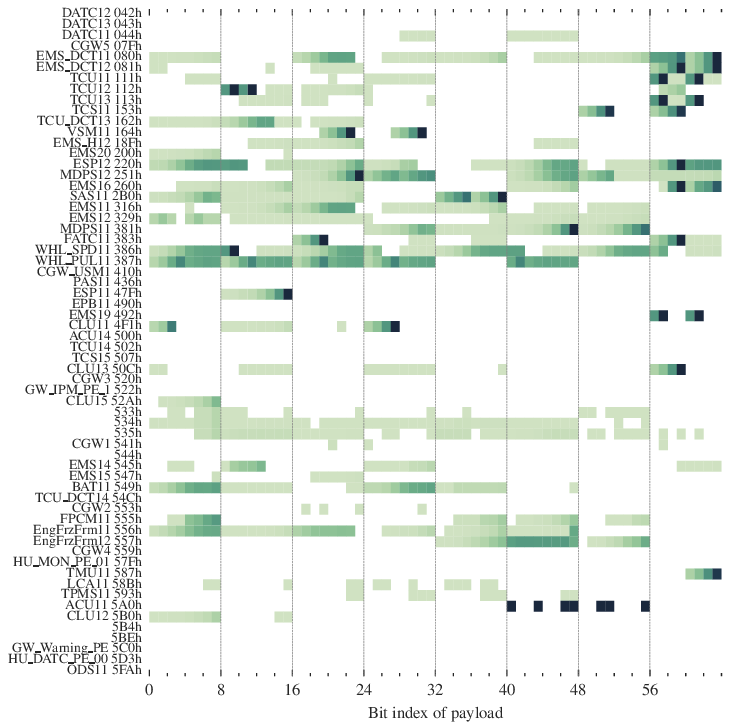}
}
\caption{Bitwise Hamming distance measurements to compare the payload dynamics of two CAN datasets captured during idling and driving. Each cell represents a value [0, 1] calculated by the number of bits flipped over the observation count. A dark cell~\legendsquare{43,27,52} means that a bit was flipped nearly every time a message arrived. A light green cell~\legendsquare{186,225,210} means that a bit was flipped only once or several times. A blank cell~\legendsquare{231,231,240} indicates no bit flips. A comparison of the two datasets reveals that the payloads changed more dynamically while the vehicle moved.}
\label{fig:hammingdistance}
\end{figure*}

\begin{equation}
  {\bf d} = \frac{1}{n-1} \sum\limits_{i=2}^{n} \left({\bf p}_i \oplus {\bf p}_{i-1}\right).
\end{equation}
where $\oplus$ is the bitwise XOR operator, ${\bf p}_i$ is the payload of the $i$-th message in a stream, and $n$ is the number of messages in a stream. For comparison, we measured the Hamming distance using the stationary dataset $\mathcal{M}_7$ and driving dataset $\mathcal{M}_1$. The measurements are shown in \autoref{fig:hammingdistance}. Note that $\mathcal{M}_7$ contains more CAN messages than $\mathcal{M}_1$. However, a higher Hamming distance can be observed for $\mathcal{M}_1$. Specifically, 968 bits were flipped once or more while the vehicle was moving, whereas only 764 bits were flipped while the vehicle was stationary. Moreover, the sum of the Hamming distance differed; $\approx 23\%$ of bit flips occurred more frequently while the vehicle was moving.

\section{Experimental Results}
\label{sec:experimental_result}
The detection performance, feasibility, and explainability of X-CANIDS are discussed in this section. First, the parameters used in the experiment are described. \autoref{table:dataset_summary} displays 62 streams and a maximum DLC of 8. Thus, we set $N=62$ and $M=64$ for the message receiver. Considering that the minimum and maximum values of the average time intervals were 0.01 s and 2 s, we initially assigned $t=0.01 {\rm s}$ and $w=200$ as baseline parameters. As noted in the previous section, there were 688 signal types. Thus, we initially achieved a $200 \times 688$-sized ${\bf S}$ every 0.01 s.

We observed that X-CANIDS did not need to examine all signals in our vehicles to detect intrusions. Many static signals (\textit{e.g.,}~the 10 signals of stream 5FAh in \autoref{table:dataset_summary}) can be inspected using a simple rule, whether or not a change occurs. Moreover, certain signals contain checksums or sequential counters that are easily predictable. We excluded static signals and signals that contained the following keywords: \textit{sum}, \textit{alive}, \textit{msgcount}, \textit{msgcnt}, \textit{paritybit}, and \textit{mul\_code}. After excluding these signals, we obtained 107 signals from the 35 streams. Thus, the final feature shape of ${\bf S}$ was $200 \times 107$.

\subsection{Parameters}
\label{subsec:parameters}

\begin{figure}[!t]
  \centering
  \includegraphics[width=\linewidth]{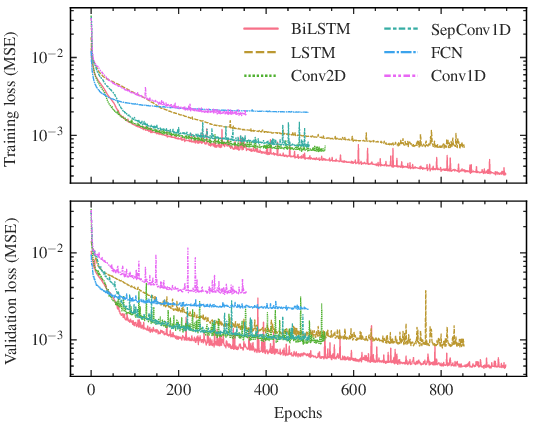}
  \caption{Learning curves of six types of autoencoders. Each training run was terminated using the early-stopping strategy. The BiLSTM-based autoencoder exhibited the smallest reconstruction error of 4.686e-4, at epoch 948.}
  \label{fig:trainloss}
\end{figure}
  
\subsubsection{Autoencoder Layer}
\label{subsec:autoencoder_layer}
First, we examine the optimal layer for ${\bf AE}$. We used six types of layers to implement the autoencoders and trained them for up to 2,000 epochs with an early-stopping patience of 50 epochs. The Adam optimizer fitted the models with a learning rate of 0.0001. A smaller MSE indicated better performance provided by a layer to ${\bf AE}$. The experimental results are presented in \autoref{fig:trainloss}. The BiLSTM layer enabled the best performance in our experiment, followed by the LSTM and Conv2D layers. 

\subsubsection{Feature Generation Parameters}
At the beginning of this section, we heuristically assigned $t=0.01{\rm s}$ and $w=200$ as baseline parameters. Two parameters can affect the detection performance and process time of X-CANIDS. Therefore, two parameters need to be chosen carefully. While both are important factors, in this section, we explore optimum values for these two parameters toward high intrusion detection performance. For the payload sampler~(\autoref{subsec:payload_sampler}), we consider the candidate time intervals as $t \in \{$0.001 s, 0.002 s, 0.005 s, 0.01 s, 0.02 s, 0.05 s, 0.1 s$\}$. For the time-series feature generator~(\autoref{subsec:time_series_feature_generator}), we consider the candidate window sizes as $w \in \{$25, 50, 75, 100, 150, 200, 300, 400$\}$. We prepared a test set with six attacks to compare the detection performances. The reader is referred to \autoref{subesc:explanation} for further information regarding the attacks. As outputs from X-CANIDS are real numbers, we use the AUC---area under the receiver operating characteristic (ROC) curve, which was rendered while adjusting the detection threshold $\Theta$---as the primary evaluation metric.

\begin{figure*}[!t]
  \centering
  \subfloat[Learning curves with sampling rate $t$]{\includegraphics[width=0.32\linewidth]{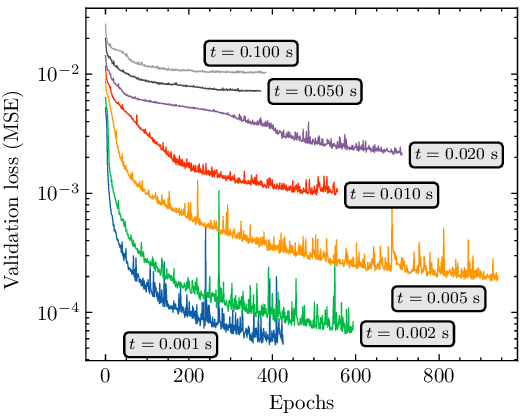}%
  \label{lc_t}} \hfil
  \subfloat[AUC with sampling rate $t$]{\includegraphics[width=0.32\linewidth]{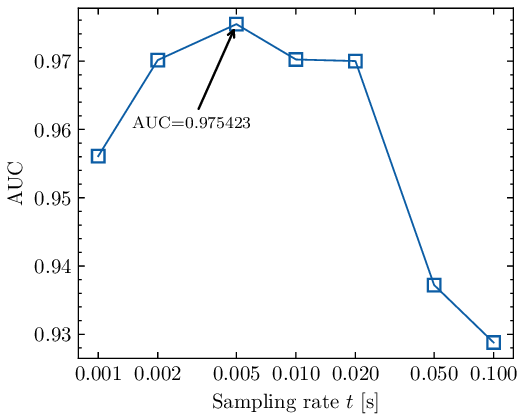}%
  \label{auc_t}} \hfil
  \subfloat[Learning curves with window size $w$]{\includegraphics[width=0.32\linewidth]{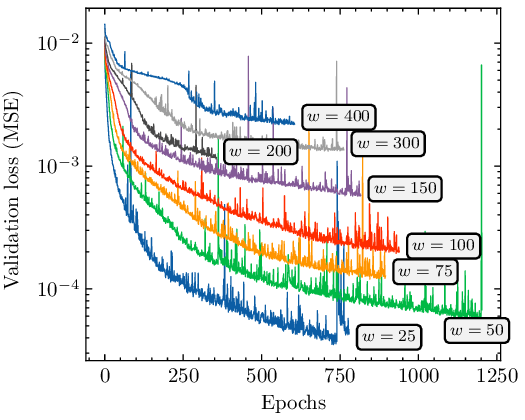}%
  \label{lc_w}} \\
  \subfloat[AUC with window size $w$]{\includegraphics[width=0.32\linewidth]{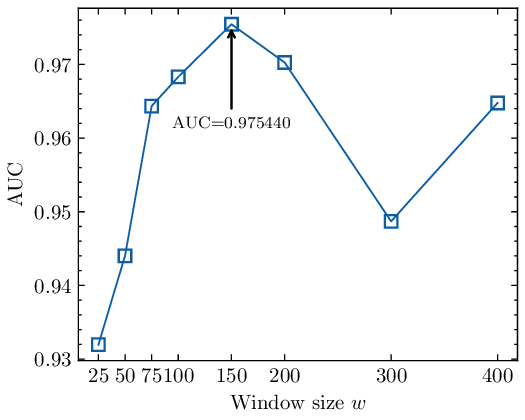}%
  \label{auc_w}} \hfil
  \subfloat[Learning curves with selected parameters]{\includegraphics[width=0.32\linewidth]{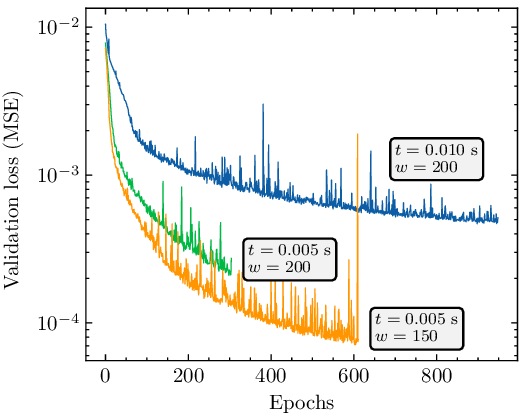}%
  \label{lc_final}} \hfil
  \subfloat[AUC with selected parameters]{\includegraphics[width=0.32\linewidth]{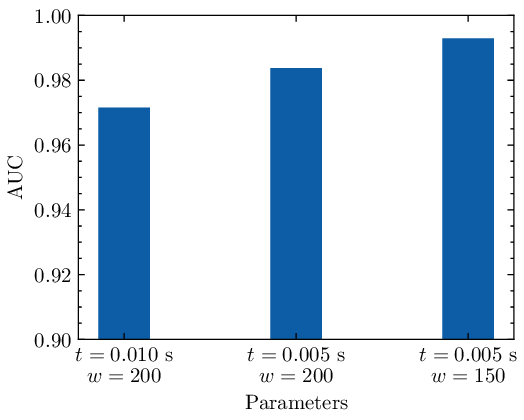}%
  \label{auc_final}} \\
  \caption{Validation losses and AUCs with parameters. In (a), (c), and (e), smaller values result in smaller validation losses. On the other hand, two parameters ($t=0.005\rm{~s}$, $w=150$) showed the best intrusion detection performances in (b) and (d), respectively. Using two values exhibited the AUC of 0.9929 in (f).}
  \label{fig:parameter_benchmark}
\end{figure*}

\autoref{fig:parameter_benchmark} shows the validation losses and intrusion detection performances with the candidate parameters. For \autoref{lc_t}--\subref*{auc_w}, we tested the candidate parameters with training sets from $\mathcal{M}_3$--$\mathcal{M}_4$. The experimental results show a clear trend---the smaller the value we choose, the smaller the reconstruction error an ${\bf AE}$ provides. However, setting a very small number for two parameters might not be a good solution because a feature ${\bf S}$ becomes representing a status of a very small time gap. Figures \autoref{auc_t} and \subref*{auc_w} support our concern that a too-small value shows a poor AUC score. Instead, in \autoref{auc_t}, we confirmed the best AUC of 0.975423 with $t=0.005\rm{~s}$. Also, in \autoref{auc_w}, we confirmed the best AUC of 0.975440 with $w=150$.

\autoref{lc_final}--\subref*{auc_final} compares two parameter combinations with the baseline parameters. In two figures, the entire training sets (\textit{i.e.,}~$\mathcal{M}_1$--$\mathcal{M}_4$) were used. Our baseline parameters ($t=0.01 {\rm s}$ and $w=200$) exhibited the moderate intrusion detection performance with the AUC of 0.9715. By changing $t$ to 0.005 from 0.010, we confirmed the better AUC of 0.9838. We also confirmed the improved AUC of 0.9929 when we change $w$ to 150 from 200 as well as $t=0.005 {\rm ~s}$. Considering X-CANIDS is an unsupervised method, the performance that we confirmed would be acceptable. Therefore, we will utilize ($t=0.005 {\rm ~s}$ and $w=150$) for the rest of the paper. 

\subsubsection{Model}
\autoref{table:ae_model} outlines the BiLSTM-based ${\bf AE}$, which was used for the remainder of the experiments. The encoder compresses a ${150 \times 107}$-sized matrix ${\bf S}$ into a $250$-sized latent vector $h$ (compression rate $\approx 1.56\%$). The decoder reconstructs ${\bf S}'$ using $h$. We assigned the parameter $\Theta=28.2$ (where $q=0.993$) for the intrusion detection.

\begin{table}[t]
  \caption{Layout of BiLSTM-based ${\bf AE}$.}
  \label{table:ae_model}
  \centering
  \begin{adjustbox}{max width=\linewidth}
  \begin{tabular}{lrrc}
  \hline
  Layer & \# parameters & Output shape & Symbol \\\hline
  Input & 0 & $150 \times 107$ & ${\bf S}$ \\
  BiLSTM & 184,040 & $150 \times 214$ & \\
  BiLSTM & 340,000 & $250$ & $h$ \\\hline
  Repeat input 150 times & 0 & $150 \times 250$ & \\
  BiLSTM & 306,448 & $150 \times 214$ & \\
  BiLSTM & 275,632 & $150 \times 214$ & \\
  Time-distributed dense & 23,005 & $150 \times 107$ & ${\bf S}'$\\\hline  
\end{tabular}
\end{adjustbox}
\end{table}

\subsection{Intrusion Detection Performance}
We tested the proposed method using the test set $\mathcal{M}_6$. We conducted attack simulations in the period 480--1440 s, half of the capture period of the test set, to obtain a label-balanced test set. We denoted the ground-truth labels as ``attack'' if a given ${\bf S}$ was affected by the attack. As each input was generated every $t=0.005 \text{~s}$, the detection result was also labeled every 0.005 s. We used precision, recall, and F1-score as the performance evaluation metrics. Our evaluation strategy was to simulate multiple attacks on the designated period and feed each dataset to X-CANIDS. Regarding the fuzzing attack, we built 17 intrusion datasets with various fuzzing rates. Regarding the fabrication, masquerade, and suspension attacks, we tried each attack on every single stream that contributed to the feature creation. To this end, we built 105 intrusion datasets. Regarding the replay attack, we built four intrusion datasets with different capture durations. Consequently, we conducted 126 experiments with different datasets.

\subsubsection{Fuzzing}

\begin{table}[t]
  \caption{Intrusion detection performance against fuzzing attacks.}
  \label{table:fuzz_detection}
  \centering
  \begin{adjustbox}{max width=\linewidth}
  \begin{tabular}{rr|rrr}
    \hline
    Fuzzing rate &  Bus load (\%) &  Precision &   Recall &  F1-score \\\hline
    10 msg./s &    100.4584 &   0.999195 & 0.911456 &  0.953311 \\
    20 msg./s &    100.9168 &   0.999251 & 0.992880 &  0.996055 \\
    30 msg./s &    101.3752 &   0.999266 & 0.998621 &  0.998943 \\
    40 msg./s &    101.8336 &   0.999261 & 0.999735 &  0.999498 \\
    50 msg./s &    102.2920 &   0.999267 & 0.999927 &  0.999597 \\
    60 msg./s &    102.7504 &   0.999262 & 0.999979 &  0.999620 \\
    70 msg./s &    103.2088 &   0.999267 & 0.999969 &  0.999618 \\
    80 msg./s &    103.6672 &   0.999194 & 0.999938 &  0.999566 \\
    90 msg./s &    104.1256 &   0.999126 & 0.999974 &  0.999550 \\
   100 msg./s &    104.5840 &   0.999121 & 0.999984 &  0.999553 \\
   200 msg./s &    109.1681 &   0.999126 & 0.999984 &  0.999555 \\
   300 msg./s &    113.7521 &   0.999121 & 0.999990 &  0.999555 \\
   400 msg./s &    118.3362 &   0.998405 & 0.999990 &  0.999197 \\
   500 msg./s &    122.9202 &   0.999023 & 0.999984 &  0.999503 \\
 1,000 msg./s &    145.8404 &   0.998322 & 0.999995 &  0.999158 \\
 1,500 msg./s &    168.7606 &   0.998322 & 0.999995 &  0.999158 \\
 2,000 msg./s &    191.6808 &   0.998322 & 0.999995 &  0.999158 \\
    \hline
    \end{tabular}
  \end{adjustbox}
  \end{table}  

Our evaluation starts with the fuzzing attack. As discussed in~\autoref{subsec:adversary}, the adversary injects CAN messages with random $a$ and ${\bf p}$. The adversary defines $a$ with one of AIDs listed in \autoref{table:dataset_summary}. We also tried various fuzzing rates. A fuzzing rate means the number of injected CAN messages per s. \autoref{table:fuzz_detection} presents the intrusion detection performance against fuzzing attacks with various fuzzing rates. The bus load column indicates the change in the number of CAN messages during the fuzzing attack. Note that the average number of CAN messages per second was $\approx 2181.48$ in the CAN bus. It can be observed that the overall detection performance was outstanding. In particular, X-CANIDS distinguished small-scale attacks with a fuzzing rate of 10 messages per second. Even when 0.4584\% of the bus load increased compared with the attack-free state, the proposed method achieved a recall of 0.911456. As the fuzzing rate increased, X-CANIDS identified nearly all intrusions, with a recall of $\approx 1$. The precision was always higher than 0.998, indicating a false positive rate of less than 0.002.

\begin{table*}[t]
  \caption{Intrusion detection performance against fabrication, masquerade, and suspension attacks.}
  \label{table:injection_detection}
  \centering
  \begin{adjustbox}{max width=\linewidth}
  \begin{tabular}{r|rrr|rrr|rrr}
    \hline
    \multicolumn{1}{l|}{\multirow{2}{*}{Target AID}} & \multicolumn{3}{l|}{Fabrication attack} & \multicolumn{3}{l|}{Masquerade attack} & \multicolumn{3}{l}{Suspension attack} \\ \cline{2-10} 
\multicolumn{1}{r|}{} & \multicolumn{1}{r}{Precision} & \multicolumn{1}{r}{Recall} & \multicolumn{1}{r|}{F1-score} & \multicolumn{1}{r}{Precision} & \multicolumn{1}{r}{Recall} & \multicolumn{1}{r|}{F1-score} & \multicolumn{1}{r}{Precision} & \multicolumn{1}{r}{Recall} & \multicolumn{1}{r}{F1-score} \\ \hline
044h &   0.755961 & 0.999993 &  0.861020 &   0.755971 & 0.999993 &  0.861027 &   0.707935 & 0.038682 &  0.073356 \\
080h &   0.999267 & 0.999979 &  0.999623 &   0.999262 & 1.000000 &  0.999631 &   0.998766 & 0.594153 &  0.745073 \\
081h &   0.999262 & 1.000000 &  0.999631 &   0.999267 & 0.999984 &  0.999625 &   0.999085 & 0.801305 &  0.889332 \\
111h &   0.999267 & 0.999979 &  0.999623 &   0.999267 & 0.999995 &  0.999631 &   0.999034 & 0.758901 &  0.862566 \\
112h &   0.999267 & 0.999979 &  0.999623 &   0.999267 & 1.000000 &  0.999633 &   0.999111 & 0.824626 &  0.903521 \\\hline
113h &   0.999267 & 0.999979 &  0.999623 &   0.999267 & 0.999995 &  0.999631 &   0.998593 & 0.520752 &  0.684531 \\
162h &   0.999267 & 0.999979 &  0.999623 &   0.999267 & 0.999995 &  0.999631 &   0.998267 & 0.422766 &  0.593981 \\
18Fh &   0.999267 & 0.999979 &  0.999623 &   0.999262 & 1.000000 &  0.999631 &   0.998264 & 0.421860 &  0.593086 \\
200h &   0.999262 & 1.000000 &  0.999631 &   0.999267 & 0.999984 &  0.999625 &   0.982244 & 0.040593 &  0.077965 \\
220h &   0.999267 & 0.999979 &  0.999623 &   0.999262 & 1.000000 &  0.999631 &   0.982360 & 0.040865 &  0.078465 \\\hline
251h &   0.999267 & 0.999922 &  0.999594 &   0.999267 & 0.999922 &  0.999594 &   0.982237 & 0.040578 &  0.077937 \\
260h &   0.999262 & 1.000000 &  0.999631 &   0.999267 & 0.999984 &  0.999625 &   0.998275 & 0.424670 &  0.595860 \\
2B0h &   0.999267 & 0.999818 &  0.999542 &   0.999267 & 0.999886 &  0.999576 &   0.982228 & 0.040557 &  0.077898 \\
316h &   0.999267 & 0.999984 &  0.999625 &   0.999256 & 1.000000 &  0.999628 &   0.999098 & 0.818234 &  0.899666 \\
329h &   0.999267 & 1.000000 &  0.999633 &   0.999262 & 1.000000 &  0.999631 &   0.999147 & 0.859115 &  0.923854 \\\hline
381h &   0.982246 & 0.040600 &  0.077976 &   0.982246 & 0.040600 &  0.077976 &   0.982246 & 0.040600 &  0.077976 \\
383h &   0.999256 & 0.999995 &  0.999625 &   0.999251 & 1.000000 &  0.999625 &   0.982235 & 0.040574 &  0.077928 \\
386h &   0.999256 & 1.000000 &  0.999628 &   0.999251 & 0.999979 &  0.999615 &   0.999044 & 0.778066 &  0.874816 \\
387h &   0.999266 & 0.999630 &  0.999448 &   0.999266 & 0.999662 &  0.999464 &   0.980727 & 0.037341 &  0.071943 \\
47Fh &   0.999266 & 0.999365 &  0.999316 &   0.999266 & 0.999521 &  0.999394 &   0.982657 & 0.041578 &  0.079781 \\\hline
4F1h &   0.999262 & 0.999979 &  0.999620 &   0.999251 & 1.000000 &  0.999625 &   0.999009 & 0.750446 &  0.857070 \\
50Ch &   0.999178 & 0.999995 &  0.999586 &   0.999267 & 0.999901 &  0.999584 &   0.982237 & 0.040584 &  0.077947 \\
52Ah &   0.999074 & 0.999792 &  0.999433 &   0.999064 & 1.000000 &  0.999532 &   0.998850 & 0.809012 &  0.893964 \\
541h &   0.999267 & 0.999938 &  0.999602 &   0.999267 & 0.999740 &  0.999503 &   0.998442 & 0.470205 &  0.639326 \\
545h &   0.999173 & 0.999995 &  0.999584 &   0.999173 & 0.999995 &  0.999584 &   0.981965 & 0.039957 &  0.076789 \\\hline
547h &   0.999178 & 0.999984 &  0.999581 &   0.999199 & 0.999979 &  0.999589 &   0.996630 & 0.217033 &  0.356444 \\
549h &   0.999173 & 0.999995 &  0.999584 &   0.999168 & 1.000000 &  0.999584 &   0.982251 & 0.040612 &  0.078000 \\
553h &   0.999266 & 0.999172 &  0.999219 &   0.999266 & 0.998751 &  0.999008 &   0.998449 & 0.472302 &  0.641263 \\
555h &   0.999168 & 0.999995 &  0.999581 &   0.999173 & 0.999990 &  0.999581 &   0.982217 & 0.040536 &  0.077859 \\
556h &   0.999168 & 1.000000 &  0.999584 &   0.999173 & 0.999984 &  0.999579 &   0.999010 & 0.835478 &  0.909955 \\\hline
557h &   0.999209 & 0.999932 &  0.999571 &   0.999173 & 0.999995 &  0.999584 &   0.986437 & 0.053375 &  0.101270 \\
58Bh &   0.999267 & 0.999984 &  0.999625 &   0.999178 & 0.999891 &  0.999534 &   0.978268 & 0.033034 &  0.063911 \\
593h &   0.999266 & 0.999755 &  0.999511 &   0.999085 & 0.999984 &  0.999534 &   0.978354 & 0.041413 &  0.079462 \\
5A0h &   0.822140 & 0.992078 &  0.899150 &   0.822323 & 0.994221 &  0.900139 &   0.748007 & 0.041658 &  0.078921 \\
5B0h &   0.749149 & 0.999951 &  0.856569 &   0.749131 & 1.000000 &  0.856575 &   0.727525 & 0.040083 &  0.075980 \\\hline
Average &   0.979597 & 0.972277 &  0.962327 &   0.979596 & 0.972341 &  0.962353 &   0.968263 & 0.328901 &  0.407648 \\\hline
\end{tabular}
\end{adjustbox}
\end{table*}

\subsubsection{Fabrication, Masquerade, and Suspension}
\autoref{table:injection_detection} presents the intrusion detection performance against the fabrication, masquerade, and suspension attacks. Regarding the fabrication and masquerade attacks, X-CANIDS showed outstanding performances with the F1-score $\geq 0.99$, except for four streams---044h, 381h, 5A0h, and 5B0h. Even though X-CANIDS successfully identified intrusions on streams 044h, 5A0h, and 5B0h with high recalls $\geq 0.992$, smaller precision confirmed that there were some false positive cases. Meanwhile, X-CANIDS did not detect intrusions on stream 381h. Based on average scores, we can rely on X-CANIDS to protect a vehicle from potential fabrication and masquerade attacks. 

Our experimental results confirmed that X-CANIDS is less effective for suspension attacks. Although we could achieved the high precision scores, small recall scores imply that X-CANIDS missed suspension attacks. The proposed method exhibited poor performance because a suspension of a certain stream does not compromise values in the buffer ${\bf P}$. As a result, ${\bf P}_{\rm copy}$ will contain legitimate values even during a suspension attack. Fortunately, considering the nature of the suspension attack, we can detect a suspension attack easily through a measurement of the number of messages in each stream. Since the precision scores are moderate, We can utilize X-CANIDS as a secondary network monitor.

\subsubsection{Replay}

\begin{table}[t]
  \caption{Intrusion detection performance against replay attacks.}
  \label{table:replay}
  \centering
  \begin{tabular}{r|rrr}
    \hline
     Capture duration (s) &  Precision &   Recall &  F1-score \\
    \hline
        0--120 &   0.999266 & 0.998954 &  0.999110 \\
      120--240 &   0.999178 & 0.999990 &  0.999584 \\
      240--360 &   0.999204 & 0.999781 &  0.999493 \\
      360--480 &   0.999262 & 0.993266 &  0.996255 \\
    \hline
  \end{tabular}
\end{table}

We conducted four further experiments to evaluate the detection performance against replay attacks. \autoref{table:replay} lists the capture durations for these experiments. For each experiment, an adversary captures a series of all legitimate CAN messages broadcasted on a CAN bus. Then, the adversary replays the series repeatedly during the designated attack period (\textit{i.e.,}~480--1440 s). For instance, in the first experiment, the content of the replayed series corresponds to a subset of the test set $\mathcal{M}_6$ in the period of 0--120 s. All streams were captured in the series. During the replay attack, the number of transmitted CAN messages per second doubled. However, the injected payloads originated from legitimate ECUs. The results in \autoref{table:replay} demonstrate that X-CANIDS is effective against replay attacks even though legitimate ECUs in our vehicle have generated all injected payloads.

\subsection{Performance Comparison with Prior Research}
As it is difficult for researchers to obtain CAN databases, previous payload-based studies~\cite{KangK16, TaylorLJ16, TariqLKW20, HossainIOFK20, LongariVZCZ20} used the raw payloads of CAN messages as inputs. For a comparison of the detection performance with prior research, we implemented a method known as CANnolo proposed by Longari \textit{et al.}~\cite{LongariVZCZ20} because the concept of the study is similar to that of our method. They proposed a self-supervised IDS using an LSTM-based autoencoder. CANnolo is supposed to be trained with benign time-series payloads in the bit representation. Consequently, we trained the model using our training set. 

\begin{figure}[!t]
  \centering
  \includegraphics[width=\linewidth]{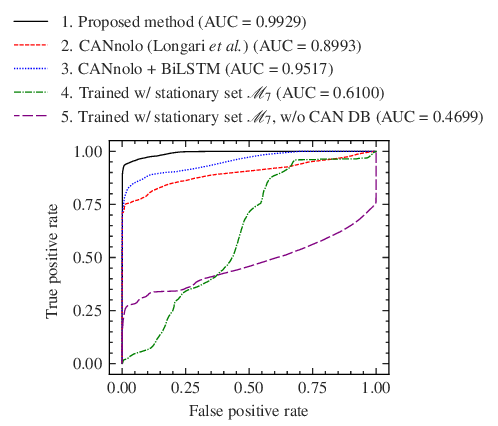}
  \caption{ROC curves. X-CANIDS (curve 1) exhibited better intrusion detection performance compared to the previous work (curve 2)~\cite{LongariVZCZ20}.}
  \label{fig:roc_curve}
\end{figure}

\autoref{fig:roc_curve} depicts five receiver operating characteristic (ROC) curves, each of which was rendered while adjusting the detection threshold $\Theta$.  Curves 1 and 2 represent the detection performances of X-CANIDS and CANnolo, respectively. X-CANIDS was superior to CANnolo, with a 0.0936 gain in the area under the curve (AUC). Furthermore, we revised CANnolo by replacing the LSTM layers with BiLSTM layers as BiLSTM exhibits a smaller reconstruction error than LSTM (see \autoref{fig:trainloss}). A comparison of curves 2 and 3 reveals that the BiLSTM layer helped CANnolo to detect anomalies more accurately. However, X-CANIDS still performed better than the revised CANnolo. We conclude that the signals that were deserialized from the raw payloads were more helpful in detecting anomalies.

\subsection{Advantages of Using Driving Dataset and Signals}
The proposed method should be trained with a driving dataset rather than a stationary or simulated dataset to ensure high detection performance. We conducted further experiments with a stationary dataset to confirm the importance of using the driving dataset in the training phase. In \autoref{fig:roc_curve}, curve 4 indicates the performance of X-CANIDS when it was trained with $\mathcal{M}_7$. A comparison of curves 1 and 4 demonstrates that the use of the driving dataset significantly improved the detection performance. 

Moreover, we assume that it is necessary to train ${\bf AE}$ with raw payloads, owing to the lack of a CAN database. Curve 5 shows the detection performance under this assumption. A comparison of curves 4 and 5 confirms that the signals aided in achieving better detection performance.

\subsection{Feasibility Consideration}
It is necessary to ensure that no bottleneck occurs owing to the computation time. The computation time is dependent on the complexity of the method and the computational power that is provided by an in-vehicle component. We selected an NVIDIA Jetson AGX Xavier, which is an automotive-grade embedded device equipped with a GPU and CAN shield, to investigate the feasibility. This device is plausible because it is used to compute automotive applications in both the vehicle industry and academia (\textit{e.g.,}~\cite{YangTQJWW23}). We implemented X-CANIDS on the device. The device could monitor the CAN bus in the vehicle using the CAN shield. The GPU allowed us to compute $f(\cdot)$ concurrently with a small batch of features. Owing to the device supporting TensorFlow natively, we could port the pretrained weights of $f(\cdot)$ from our PC.

\subsubsection{Throughput}
We first measure the throughput of our ${\bf AE}$ on the device. Note that we chose $t=0.005 {\rm ~s}$. It means that the feature generator generates 200 features per s. Consequently, the throughput must be equal to or greater than 200 samples per s. We tried 11 batch sizes and summarized the result in \autoref{table:inference_time_gpu}. We denote the batch size as $B$. We confirmed that the device can process up to $\approx 2,010$ samples per s when we use $B=1024$. In order to minimize the batch completion time as well as the detection latency, we decide the $B=8$. 

\newcommand{\specialcell}[2][r]{\begin{tabular}[#1]{@{}c@{}}#2\end{tabular}}
\begin{table}[t]
  \caption{Throughput and inference time on NVIDIA Jetson AGX Xavier.}
  \centering
  \label{table:inference_time_gpu}
  \begin{adjustbox}{max width=\linewidth}
  \begin{tabular}{rrr}
  \hline
  Batch size $B$ & \specialcell{Throughput\\(samples/s)} & \specialcell{Inference time $t_\beta$\\(ms/sample)} \\\hline
4     &   114.6033  & 8.7258  \\
8     &   239.1564  & 4.1814  \\
16    &   476.0859  & 2.1005  \\
32    &   870.1233  & 1.1493  \\
64    & 1,257.3287  & 0.7953  \\
128   & 1,526.7854  & 0.6550  \\
256   & 1,773.3477  & 0.5639  \\
512   & 1,928.1721  & 0.5186  \\
1,024 & 2,010.5995  & 0.4974  \\
2,048 & 1,881.2511  & 0.5316  \\
4,096 & 1,234.2686  & 0.8102  \\\hline
  \end{tabular}
\end{adjustbox}
\end{table}

\subsubsection{Detection latency}
The detection latency is the time gap between the attack and alert. When $B=8$ was selected, the detection latency was derived as follows:
\begin{equation}
  \begin{aligned}
  &\text{Detection latency} = zt + t_\alpha + Bt_\beta
  \\&= z\cdot5\text{~ms} + 4.8 \text{~ms} + 8\cdot4.1814\text{~ms}
  \\&= \begin{cases}
      38.2512    \text{~ms}, & \text{if $z=0$, the batch is full on arrival.}\\
      73.2512 \text{~ms}, & \text{if $z=7$, the batch is empty.}\\
    \end{cases}
  \end{aligned}
\end{equation}
In the above, $t_\alpha=4.8 \text{~ms}$ stands for the time consumption that the feature generator takes, and $t_\beta$ stands for the inference time per sample (see \autoref{table:inference_time_gpu}). Also, $z\in\{0..(B-1)\}$ is the number of inputs required to complete a batch; thus, $zt$ represents the batch completion time. We conclude that X-CANIDS provides an intrusion alert with a deterministic latency of no greater than 73.2512~ms.

The deterministic latency is high compared to the minimum time intervals of existing streams in our vehicle (\textit{i.e.,}~10~ms, see~\autoref{table:dataset_summary}). Carmakers considering installing an intrusion response system (IRS) should note that while X-CANIDS can be used for intrusion detection, it cannot be seamlessly integrated with their IRS to mitigate identified cyberattacks in time without further optimizing the detection latency. We have deferred the optimization to future work.

\subsubsection{CPU, RAM and GPU usage}
We utilized the embedded device only to evaluate the feasibility of X-CANIDS. However, the device may run other processes along with X-CANIDS in a real-world scenario. For those who are considering the scenario, we measured the CPU and RAM usage on our device. Our inference program---implemented using C++ with standard template library---takes the CPU utilization of $\approx$150\% (100\% per core, 8 cores total) and the memory space of $\approx$2.7GB including deep learning libraries. We expect the CPU and RAM requirements steady in other environments because X-CANIDS does not rely on sparse data (\textit{e.g.,}~one-hot vector) or dynamic memory allocation. The bus load of a CAN-based IVN does not affect the performance either because the feature generator builds an ${\bf S}$ every designated $t$ s. We had no choice but to use two threads since $t_\alpha=4.8 \text{~ms}$ was nearly equal to the feature creation interval $t$; one thread generates features, and another thread deals with the rest---the batch compilation, feeding a batch to the model, and measure the error rate. Finally, the GPU utilization was measured as $\approx$63\% on average when $B=8$.

\subsection{Explanation of Detection Results}
\label{subesc:explanation}
\begin{figure*}[!t]
\centering
\begin{adjustbox}{max width=\linewidth}
\includegraphics[width=\linewidth]{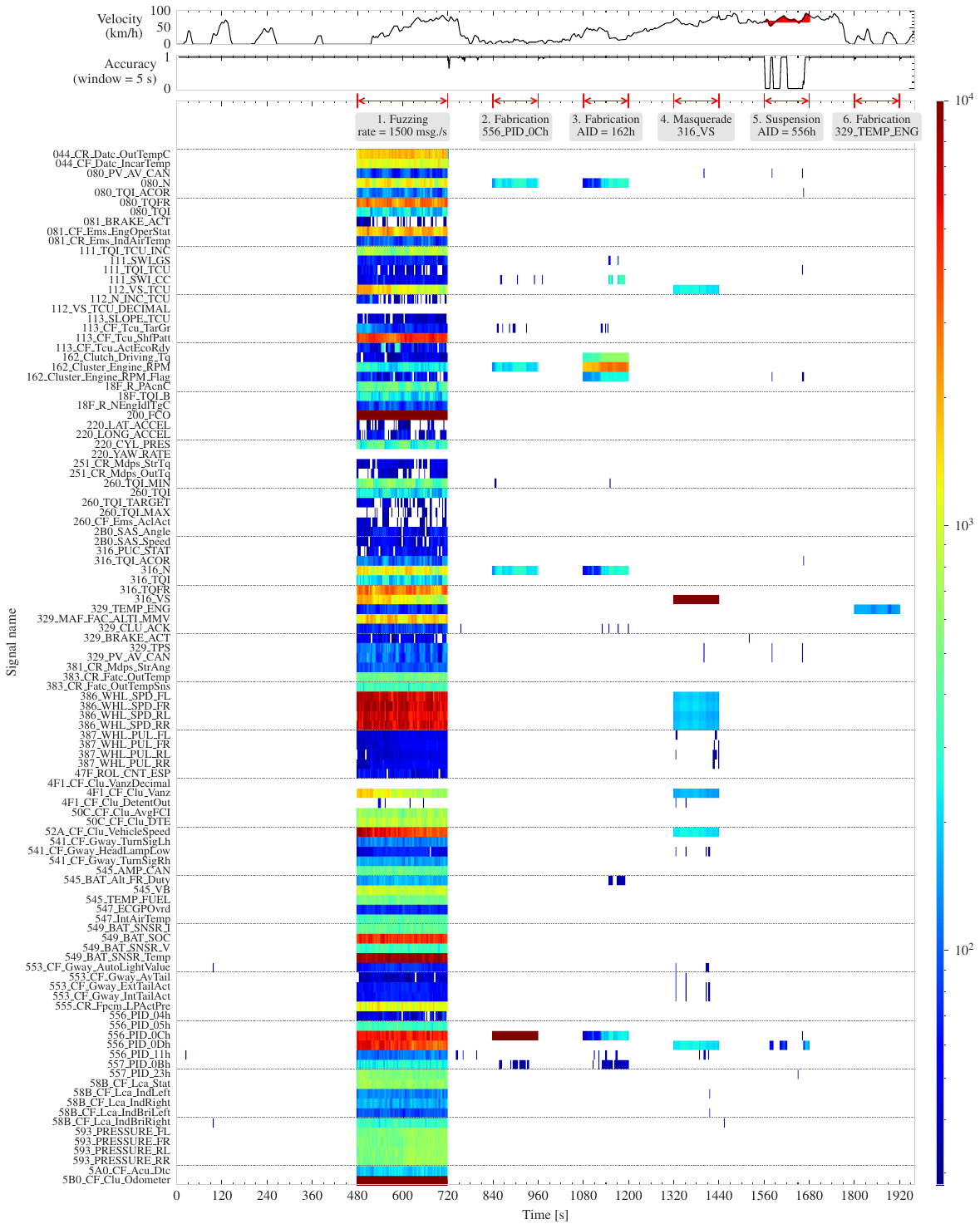}
\end{adjustbox}
\caption{Error rates over test dataset $\mathcal{M}_6$. The error rates are depicted as $28.2 = \Theta \le$ blue $\le 10^2 \le$ green $\le 10^3 \le$ red $\le 10^4+$. We conducted attacks in six periods. In attack period 1, the error rates were exceeded on many signals owing to intensive fuzzing. In attack periods 2--6, we could distinguish the exact target signals showing a maximum error rate at a certain time. The reader is referred to the color version of this page for interpretation of the figure.
}
\label{fig:signal_prediction}
\end{figure*}

In this section, we discuss the explainability of our proposed framework. A heatmap that plots the error rates of the 107 signals over time is presented in \autoref{fig:signal_prediction}. We conducted six attacks on our test set and measured their error rates. The figure shows only the error rates that exceeded the detection threshold $\Theta=28.2$. That is, the marked points in the figure represent the predicted intrusions. The attack period and description are indicated at the top of the heatmap. The moving average of the detection accuracy is also provided.

\subsubsection{Period 1---Fuzzing} An intensive fuzzing attack was conducted during 480--720 s and the error rate exceeded the detection threshold for almost all signals. The $\rm{argmax}({\bf r})$ function continuously pointed out signal 5B0\_CF\_Clu\_Odometer. Nevertheless, an expert who is responsible for incident response would be able to identify the type of attack as fuzzing because of the multiple simultaneous errors. Unfortunately, during this period, the proposed framework did not work for the following signals: 112\_VS\_TCU\_DECIMAL, 220\_YAW\_RATE, and 4F1\_CF\_Clu\_VanzDecimal.

\subsubsection{Period 2---Fabrication} We considered an adversary who attempts to fabricate a portion of the payloads for signal 556\_PID\_0Ch during 840--960 s. The signal is part of the OBD-II freeze frame containing the parameter ID 0Ch. That is, the signal represents the current engine RPM~\cite{obd_ii_pids}. The experimental results showed that 556\_PID\_0Ch reached the highest error rate of $\approx 10^4$ during this period. Three signals representing the RPM, namely 080\_N, 162\_Cluster\_Driving\_RPM, and 316\_N, also exhibited high error rates.

\subsubsection{Period 3---Fabrication} An adversary who injects CAN messages with $a$=162h was assumed during 1080--1200 s. We can see that three signals belonging to stream 162h exhibited high error rates. In particular, $\rm{argmax}({\bf r})$ successfully pointed out signal 162\_Cluster\_Enging\_RPM.

\subsubsection{Period 4---Masquerade} We considered an adversary who attempts to report a fraudulent current velocity to the driver via an instrumental cluster. For this purpose, we simulated a masquerade attack that changes signal 316\_VS during 1320--1440 s. As shown in the figure, the signal exhibited the highest error rate. Furthermore, other speed-related signals exhibited high error rates simultaneously, including the four wheel speed signals that are defined in stream 386h. 

\subsubsection{Period 5---Suspension} We performed a suspension attack for stream 556h during 1560--1680 s. As demonstrated in \autoref{table:injection_detection}, the proposed method is ineffective against suspension attacks. The figure also supports the weakness of X-CANIDS. We observed that the detection accuracy increased and decreased during this period. The proposed framework identified anomalies in signal 556\_PID\_0Dh, which represents the current velocity (see the description of parameter ID 0Dh in \cite{obd_ii_pids}). The error rate of the signal exceeded the detection threshold when the vehicle velocity differed from 68 km/h (cf., the red-filled area in the velocity chart and change in the moving accuracy).

\subsubsection{Period 6---Fabrication} Finally, motivated by previous works~\cite{ShahriarXMLH22, VermaIBHMKC22}, we considered a max coolant temperature attack. In our dataset, signal 329\_TEMP\_ENG deals with the engine coolant temperature. Thus, we conducted a fabrication attack on the signal. It can be observed that $\rm{argmax}({\bf r})$ successfully pointed out the exact target signal during this period.

\section{Related Works}
\label{sec:related_work}

\begin{table}[t]
\caption{Previous work that proposed in-vehicle IDSs for CAN buses.}
\label{table:related_work}
\setlength{\tabcolsep}{4pt} 
\centering
\begin{tabular}{lcccclr}
\hline
Research & \rot{Timestamp} & \rot{AID sequence} & \rot{Payload} & \rot{Signal} & Output & \rot{Inference time~~} \\ \hline
Müter and Asaj~\cite{MuterA11} & \checkmark &  & \checkmark &  & Binary & --- \\
Kang and Kang~\cite{KangK16} &  &  & \checkmark &  & Binary & 2--5 ms$^\dagger$ \\
Marchetti and Stabili~\cite{MarchettiS17} &  & \checkmark &  &  & Binary & --- \\
Taylor \textit{et al.}~\cite{TaylorJL15} & \checkmark &  & \checkmark &  & Binary & --- \\
Song \textit{et al.}~\cite{SongKK16} & \checkmark &  &  &  & Binary & 1 ms$^\dagger$ \\
Taylor \textit{et al.}~\cite{TaylorLJ16} &  &  & \checkmark &  & Binary & --- \\
Marchetti \textit{et al.}~\cite{MarchettiSGC16} & \checkmark &  &  &  & Binary & --- \\
Stabili \textit{et al.}~\cite{StabiliMC17} &  &  & \checkmark &  & Binary & --- \\
Markovitz and Wool~\cite{MarkovitzW17} &  &  &  & \checkmark & Binary & --- \\
Wasicek \textit{et al.}~\cite{WasicekPWBS17} &  &  &  & \checkmark & Real number & --- \\
Tomlinson \textit{et al.}~\cite{TomlinsonBSK18} & \checkmark &  &  &  & Binary & --- \\
Olufowobi \textit{et al.}~\cite{OlufowobiYZB19} & \checkmark &  &  &  & Binary & 9--10 ms$^\dagger$ \\
Young \textit{et al.}~\cite{YoungOBZ19} & \checkmark &  &  &  & Binary & --- \\
Katragadda \textit{et al.}~\cite{KatragaddaDRG20} &  & \checkmark &  &  & Binary & 151 ms$^\ddagger$ \\
Song \textit{et al.}~\cite{SongWK20} &  & \checkmark &  &  & Binary & 5--6.7 ms$^\dagger$ \\
Longari \textit{et al.}~\cite{LongariVZCZ20} &  &  & \checkmark &  & Binary & --- \\
Hossain \textit{et al.}~\cite{HossainIOFK20} &  & \checkmark & \checkmark &  & Category & --- \\
Tariq \textit{et al.}~\cite{TariqLKW20} & \checkmark & \checkmark & \checkmark &  & Category & 14--73 ms$^\dagger$ \\
Song and Kim~\cite{SongKim21} &  & \checkmark &  &  & Binary & --- \\
Shahriar \textit{et al.}~\cite{ShahriarXMLH22} &  &  &  & \checkmark & Real number & --- \\
Hoang and Kim~\cite{HoangK22} &  & \checkmark &  &  & Binary & 0.63 ms$^\dagger$ \\
X-CANIDS (this work) &  &  &  & \checkmark & Real number & 4.18 ms$^{\ddagger}_*$ \\ \hline
\end{tabular}
\\[.075in]
\begin{minipage}[t]{\linewidth}
$^\dagger$per CAN message \hspace{.075in} $^\ddagger$per feature \hspace{.075in} $^*$measured on an embedded device
\end{minipage}
\end{table}

In this section, related works on in-vehicle IDSs for CAN buses are reviewed. \autoref{table:related_work} lists 21 previous studies, with the input type used to detect the intrusion, output type, and inference time presented in the respective paper. In many cases, an IDS returns a binary for each input to indicate whether the vehicle is attacked. Meanwhile, two studies~\cite{HossainIOFK20, TariqLKW20} proposed IDSs that return a categorical value. This category refers to the type of attack. These IDSs require labeled training sets to recognize the attack type. Two signal-aware IDSs (\cite{WasicekPWBS17, ShahriarXMLH22}) return a real number that can be used as the detection result and confidence score.

The primary goal of an in-vehicle IDS is to detect anomalies effectively. Moreover, an IDS should be implemented on an ECU or embedded device that provides limited computational power. Therefore, IDSs should be tested on these devices to confirm their feasibility. In particular, the inference time should be compared with the feature-creation frequency. However, excluding this work, only seven studies measured the inference time. Moreover, none of the studies mentioned that the inference time was measured using an embedded device.

An in-vehicle IDS assesses the in-vehicle traffic using the timestamp, AID sequence, payload, and/or signal. In the remainder of this section, we categorize the related studies according to the input type.

\subsection{Time Interval-Based IDS}
A timestamp is not an officially supported field in the CAN frame. Nevertheless, an ECU can measure the relative time of the transmission itself because every ECU that shares the same medium is time synchronized bit by bit. ECUs tend to report their status periodically. Time-interval-based IDSs exploit these mechanisms. The measurement of timestamps is an efficient means of detecting anomalies. This also enables the detected intrusions to be understood. In contrast, a time-interval-based IDS does not work for the sporadic transmission of CAN messages.

Most early studies leveraged the periodic transmission mechanism. For example, in 2011, Müter and Asaj~\cite{MuterA11} explored the applicability of the entropy-based anomaly detection method. The proposed method takes advantage of the periodicity of IVN traffic. They measured the normal probability distribution of attack-free CAN data and compared it with that of abnormal CAN data. However, the study lacked an explicit threshold determination method or evaluation result, such as the detection accuracy. In 2016, Song \textit{et al.}~\cite{SongKK16} proposed a rule-based intrusion detection method that measures the time interval of two adjacent CAN messages. When the time interval of a new CAN message is shorter than the threshold, the proposed IDS considers the message as an intrusion. The experimental results demonstrated high detection performance with handcrafted thresholds. However, the method for determining the threshold remains to be improved. Marchetti \textit{et al.}~\cite{MarchettiSGC16} proposed an entropy-based anomaly detector for CAN buses. Tomlinson \textit{et al.}~\cite{TomlinsonBSK18} adopted the ARIMA model and Z-score to identify CAN message timing anomalies in a time window. Olufowobi \textit{et al.}~\cite{OlufowobiYZB19} proposed an intrusion detection method known as SAIDuCANT, which leverages the periodic message behavior in the CAN bus. Young \textit{et al.}~\cite{YoungOBZ19} adopted  fast Fourier transform to measure CAN message update frequency in a stream.

\subsection{Sequence-Based IDSs}
Sequence prediction is a well-known machine-learning problem. A sequence predictor predicts the next symbol based on previously observed categorical data. A sequence of AIDs can be used as sequence predictors to model the attack-free state of IVNs for in-vehicle traffic monitoring. Sequence-based IDSs can be used in all types of CAN-based IVNs because it is easy to compile AID sequences. However, these methods exhibit several drawbacks. Particularly, it is difficult to understand why a given sequence is classified as abnormal. Moreover, an adversary may falsify a legitimate sequence by conducting an intensive replay attack.

Marchetti and Stabili~\cite{MarchettiS17} compiled a transition matrix to represent the recurring patterns of two adjacent AIDs. The transition matrix is used to evaluate the CAN bus stream in the inference phase. Katragadda \textit{et al.}~\cite{KatragaddaDRG20} proposed a message sequence-based anomaly detection method that builds a frequent sequence tree, where each node represents a subsequence and each edge measures an observation count. 
The sequence length should be carefully selected because detection performance and feasibility are dependent thereon. Tariq \textit{et al.}~\cite{TariqLKW20} combined the benefits of rule-based models and neural networks. They claimed that their heuristic model works for known attack signatures, whereas neural networks can cope with unknown attacks. This is the only work that simultaneously examined the timestamp, payload, and AID sequences. Indeed, their method requires a substantially longer inference time than those of other approaches. Song \textit{et al.}~\cite{SongWK20} proposed a CNN that is a variation of Inception-ResNet to examine AID sequences. Their detection model outperformed conventional detection methods. However, a limitation of the proposed method is that it requires a labeled dataset to train the model. To overcome this drawback, they proposed another training method~\cite{SongKim21} to train their CNN model in an unsupervised manner. Hoang and Kim~\cite{HoangK22} proposed an adversarial autoencoder that attempts to reconstruct $29\times29$-sized features, which represent 29 continuous AIDs in bits.

\subsection{Payload-Based IDSs}
Payload-based IDSs evaluate the payloads in CAN messages, where each payload is represented as bit sequences. The training set needs to be carefully prepared to use payload-based IDSs in real-world scenarios; otherwise, the payload dynamics could cause false alarms. 

Taylor \textit{et al.}~\cite{TaylorJL15} measured the number of packets and average Hamming distance of the CAN message payloads in a sliding window. The statistical features that were derived from these two values were used to train a one-class support vector machine. Their experimental results were dependent on the window size for feature generation. Unfortunately, the unsupervised method exhibited a considerable false positive rate with the incorrect window size. Kang and Kang~\cite{KangK16} proposed a binary intrusion detection method based on a fully connected neural network, which uses 64-dimensional bit sequences (\textit{i.e.,}~a payload of CAN messages) and then returns a logistic value of 0 or 1. They used a packet generator known as OCTANE to evaluate the proposed method. Despite the high performance of the experimental results, this work exhibits two limitations: (1) the method was evaluated using only three streams and (2) the simulated payloads that were used in the experiment may not reflect real-world situations. Taylor \textit{et al.}~\cite{TaylorLJ16} proposed an intrusion detector that consists of LSTM-based models for each stream. Each model uses a $20\times64$-sized matrix that comprises 20 subsequences of CAN payloads in the bit representation. Subsequently, the model predicts the next payload of a stream. They considered five loss metrics to measure the anomaly score. Stabili \textit{et al.}~\cite{StabiliMC17} measured the Hamming distance for each stream. They classified each stream into the no distance, small distance, and mid-distance ranges. When a stream exceeds a given distance range in the inference phase, it is classified as an anomaly. However, the method was evaluated using only fuzzing attacks. Hossain \textit{et al.}~\cite{HossainIOFK20} developed a supervised IDS using an LSTM model that considers the raw payloads of a CAN message. Longari \textit{et al.}~\cite{LongariVZCZ20} deployed an anomaly detection system based on LSTM autoencoders. They implemented an LSTM model for each stream to reconstruct a time-series bit sequence of payloads.

\subsection{Signal-Aware IDSs}




To the best of our knowledge, Markovitz and Wool~\cite{MarkovitzW17} were the first to use portions of the payload in CAN messages to detect intrusions, as opposed to using an entire payload. They divided a 64-bit payload into several fields, each of which was assigned as a constant, categorical, counter, or sensor type. A rule-based detection algorithm was considered for each data type. Wasicek \textit{et al.}~\cite{WasicekPWBS17} proposed an IDS that uses 54 types of signals that are obtained via OBD-II PIDs. They reported that a fully connected network with a bottleneck can be used to measure the anomaly score (\textit{i.e.,}~the reconstruction error). In 2022, Shahriar \textit{et al.}~\cite{ShahriarXMLH22} proposed a signal-aware IDS, named CANShield. A 2D CNN was designed to reconstruct the signals that were deserialized from raw payloads. They used CAN-D~\cite{VermaBSHI21}, which is an automatic dissector for CAN traffic, to obtain these signals. 

\subsection{Comparison with related works}
In the literature, two related works~\cite{WasicekPWBS17, ShahriarXMLH22} and X-CANIDS are closely aligned in that \textit{``an autoencoder takes signals to detect in-vehicle intrusion.''} In this section, we provide a comparative analysis to discuss the advantages of X-CANIDS in terms of signal processing and the detection model.

\BfPara{Signal process}
The method in~\cite{WasicekPWBS17} takes signals from OBD-II responses, while X-CANIDS and CANShield~\cite{ShahriarXMLH22} use signals from CAN messages. OBD-II responses are an alternative source to obtain signals without a CAN database. However, these signals primarily reflect powertrain-related sensor values, as the main objective of OBD-II is vehicle diagnosis. As a result, the method in~\cite{WasicekPWBS17} is insufficient to protect non-powertrain applications. Another limitation is that an OBD-II response will yield compromised signals once the powertrain has been compromised. Therefore, signals from CAN messages are more beneficial as they can reflect attempted attacks instantly. Meanwhile, CANShield has two limitations in signal processing. First, the method only analyzes a few pre-selected signals while ignoring the rest by design. Second, the order of signals can affect learning efficacy since the signals are converted into a 2D image. Compared to these related works, X-CANIDS has the following advantages: (1) it uses signals from CAN messages, (2) it can analyze all available signals, and (3) it ensures robust learning efficacy regardless of the order of signals.

\BfPara{Model}
The methods in~\cite{WasicekPWBS17, ShahriarXMLH22} use an FCN- and a Conv2D-based autoencoder, respectively. In addition to these types of layers, we have tested six different layers to implement autoencoders. Our experimental results, using real driving datasets, show that the BiLSTM layer is superior to the FCN and Conv2D layers for the signal reconstruction problem (refer to \autoref{subsec:autoencoder_layer} and \autoref{fig:trainloss}). Investigating the most effective layer is also a key contribution of this work.

\section{Discussion and Conclusions}
\label{sec:conclusion}

\subsection{Limitation}
We here discuss three limitations as well as remediation strategies for them. First, we used only one vehicle to evaluate X-CANIDS. We found it difficult to arrange another car owing to the lack of a CAN database. To tackle the issue, we encourage carmakers to share their CAN databases with academia for research purposes. Otherwise, providing an application programming interface that allows researchers to obtain signals can be an alternative solution. 

Second, X-CANIDS is insufficient for detecting suspension attacks. To this end, our future work will expand the proposed method to examine time-interval-based features along with signal features. Except for the masquerade attack, the rest of the attacks can be conducted by injecting or suppressing CAN messages. Therefore, analyzing a transmission period per stream would be a promising approach. The transmission period can also be considered an intuitive explanation because an expert can easily compare a current one with an expected one.

Finally, X-CANIDS takes some time to process the input and raise the alarm. Many streams have an update interval of 10~ms. However, we confirmed the detection latency of 73.2512~ms in the worst case. It implies there is a time gap exposed to the attack without awareness. To tackle the issue, we could get prompt detection results by considering a lightweight autoencoder or dedicated hardware that accelerates the autoencoder. IDSs with substantial computational demands may not be appropriate for all vehicle tiers, from low- to high-end. To achieve real-time detection at a reasonable cost, there is a compelling need for more lightweight IDS mechanisms.

\subsection{Remarks and Conclusion}
Recent vehicles are driven by software and have a broad attack surface. So far, many studies have been proposed for the precise detection of intrusions on in-vehicle networks. However, due to the lack of information on payload serialization, only a few studies have considered analyzing signals of CAN messages. Also, feasibility considerations are lacking. In response, cybersecurity regulations, including UNR 155, enforce the installation of an IDS inside vehicles and the analysis of cyberattacks. Therefore, the feasibility and explainability of in-vehicle IDSs are important for vehicle industries.

In this study, we have proposed X-CANIDS, in which the feature generator is designed to process live streams and create a time-series representation of the signals. A CAN database is combined with X-CANIDS to deserialize the signals from the CAN message payloads. We tested six types of autoencoders. The LSTM layer has often been considered in the literature \cite{TaylorLJ16, TariqLKW20, HossainIOFK20, LongariVZCZ20}. A Conv2D-based autoencoder has also been employed to model signals~\cite{ShahriarXMLH22}. However, we demonstrated that the BiLSTM-based autoencoder outperformed the LSTM and 2D-CNN autoencoders. Then, we explored two parameters regarding feature generation. X-CANIDS expects an onboard AI-inference device to leverage the model. In case of a lack of such a device, a driver may consider installing a device like an after market autonomous driving kit of Comma.ai.

In summary, the experimental results suggest that X-CANIDS detects zero-day intrusions that are not observed during the training phase. In particular, the proposed method offers an advantage for masquerade attacks that cannot be detected by time-interval- or sequence-based IDSs. X-CANIDS also offers outstanding performance against fuzzing, fabrication, and replay attacks.

We have considered the feasibility and explainability. To the best of our knowledge, these characteristics have not been considered in the previous works for CAN IDSs. Our feasibility evaluation confirms that X-CANIDS is able to be implemented on an embedded device to monitor live in-vehicle traffic while driving. The explainability provides an explicit hint about target ECUs or compromised signals. The explainability of X-CANIDS will help incident response teams analyze conducted cyberattacks. For that, X-CANIDS requires a CAN database. We claim that our method will be valuable to all carmakers because they can access CAN databases for their vehicles. 

\bibliographystyle{IEEEtran}
\bibliography{bibliography.bib}

\begin{thebibliography}{10}
\providecommand{\url}[1]{#1}
\csname url@samestyle\endcsname
\providecommand{\newblock}{\relax}
\providecommand{\bibinfo}[2]{#2}
\providecommand{\BIBentrySTDinterwordspacing}{\spaceskip=0pt\relax}
\providecommand{\BIBentryALTinterwordstretchfactor}{4}
\providecommand{\BIBentryALTinterwordspacing}{\spaceskip=\fontdimen2\font plus
\BIBentryALTinterwordstretchfactor\fontdimen3\font minus \fontdimen4\font\relax}
\providecommand{\BIBforeignlanguage}[2]{{%
\expandafter\ifx\csname l@#1\endcsname\relax
\typeout{** WARNING: IEEEtran.bst: No hyphenation pattern has been}%
\typeout{** loaded for the language `#1'. Using the pattern for}%
\typeout{** the default language instead.}%
\else
\language=\csname l@#1\endcsname
\fi
#2}}
\providecommand{\BIBdecl}{\relax}
\BIBdecl

\bibitem{Cho2016a}
K.-T. Cho and K.~G. Shin, ``Error handling of in-vehicle networks makes them vulnerable,'' in \emph{Proc. ACM CCS '16}, 2016, pp. 1044--1055.

\bibitem{LeeJK17}
H.~Lee, S.~H. Jeong, and H.~K. Kim, ``{OTIDS}: A novel intrusion detection system for in-vehicle network by using remote frame,'' in \emph{Proc. PST '17}, 2017.

\bibitem{MillerV15}
C.~Miller and C.~Valasek, ``Remote exploitation of an unaltered passenger vehicle,'' in \emph{Proc. Black Hat USA '15}, Aug. 2015, pp. 1--91.

\bibitem{KimKJPK21}
K.~Kim, J.~S. Kim, S.~Jeong, J.-H. Park, and H.~K. Kim, ``Cybersecurity for autonomous vehicles: Review of attacks and defense,'' \emph{Computers \& Security}, vol. 103, p. 102150, 2021.

\bibitem{JoCNWL17}
H.~J. Jo, W.~Choi, S.~Y. Na, S.~Woo, and D.~H. Lee, ``Vulnerabilities of {Android} {OS}-based telematics system,'' \emph{Wireless Personal Communications}, vol.~92, no.~4, pp. 1511--1530, 2017.

\bibitem{Gayou18}
S.~Gayou, ``Jailbreaking {Subaru StarLink},'' \url{https://github.com/sgayou/subaru-starlink-research}, Nov. 2018.

\bibitem{TeamFluoroacetate19}
``Tesla car hacked at {Pwn2Own} contest,'' \url{https://www.zdnet.com/article/tesla-car-hacked-at-pwn2own-contest/}, Mar. 2019.

\bibitem{CostantinoM19}
G.~Costantino and I.~Matteucci, ``{CANDY CREAM} - hacking infotainment {Android} systems to command instrument cluster via {CAN} data frame,'' in \emph{Proc. IEEE CSE '19}, 2019, pp. 476--481.

\bibitem{TencentMBUX21}
{Tencent Keen Security Lab}, ``{Mercedes-Benz MBUX} security research report,'' \url{https://keenlab.tencent.com/en/whitepapers/Mercedes_Benz_Security_Research_Report_Final.pdf}, Tech. Rep., May 2021.

\bibitem{JoC21}
H.~J. Jo and W.~Choi, ``A survey of attacks on {Controller Area Networks} and corresponding countermeasures,'' \emph{IEEE Transactions on Intelligent Transportation Systems}, pp. 6123--6141, Jul. 2022.

\bibitem{UNR155}
``{UN Regulation No. 155} - cyber security and cyber security management system,'' E/ECE/TRANS/505/Rev.3/Add.154, Apr. 2021.

\bibitem{HossainIOFK20}
M.~D. Hossain, H.~Inoue, H.~Ochiai, D.~Fall, and Y.~Kadobayashi, ``{LSTM}-based intrusion detection system for in-vehicle {CAN} bus communications,'' \emph{IEEE Access}, vol.~8, pp. 185\,489--185\,502, 2020.

\bibitem{TariqLKW20}
S.~Tariq, S.~Lee, H.~K. Kim, and S.~S. Woo, ``{CAN-ADF}: The {Controller Area Network} attack detection framework,'' \emph{Computers \& Security}, vol.~94, p. 101857, 2020.

\bibitem{WasicekPWBS17}
A.~Wasicek, M.~D. Pes{\'e}, A.~Weimerskirch, Y.~Burakova, and K.~Singh, ``Context-aware intrusion detection in automotive control systems,'' in \emph{Proc. 5th ESCAR USA '17}, 2017, pp. 21--22.

\bibitem{ShahriarXMLH22}
\BIBentryALTinterwordspacing
M.~H. Shahriar, Y.~Xiao, P.~Moriano, W.~Lou, and Y.~T. Hou, ``{CANShield}: Signal-based intrusion detection for {Controller Area Networks},'' 2022. [Online]. Available: \url{https://arxiv.org/abs/2205.01306}
\BIBentrySTDinterwordspacing

\bibitem{VermaBSHI21}
M.~E. Verma, R.~A. Bridges, J.~J. Sosnowski, S.~C. Hollifield, and M.~D. Iannacone, ``{CAN-D}: A modular four-step pipeline for comprehensively decoding {Controller Area Network} data,'' \emph{IEEE Transactions on Vehicular Technology}, vol.~70, no.~10, pp. 9685--9700, 2021.

\bibitem{YoungSZ20}
C.~Young, J.~Svoboda, and J.~Zambreno, ``Towards reverse engineering {Controller Area Network} messages using machine learning,'' in \emph{Proc. IEEE WF-IoT '20}, 2020, pp. 1--6.

\bibitem{PeseSCNCS19}
M.~D. Pes\'{e}, T.~Stacer, C.~A. Campos, E.~Newberry, D.~Chen, and K.~G. Shin, ``{LibreCAN}: Automated {CAN} message translator,'' in \emph{Proc. ACM CCS '19}, ser. CCS '19.\hskip 1em plus 0.5em minus 0.4em\relax New York, NY, USA: Association for Computing Machinery, 2019, pp. 2283--2300.

\bibitem{KangSJK18}
T.~U. Kang, H.~M. Song, S.~Jeong, and H.~K. Kim, ``Automated reverse engineering and attack for {CAN} using {OBD-II},'' in \emph{Proc. IEEE VTC-Fall '18}, 2018, pp. 1--7.

\bibitem{VermaBH18}
M.~Verma, R.~Bridges, and S.~Hollifield, ``Actt: Automotive {CAN} tokenization and translation,'' in \emph{Proc. CSCI '18}, 2018, pp. 278--283.

\bibitem{HuybrechtsVBBH17}
T.~Huybrechts, Y.~Vanommeslaeghe, D.~Blontrock, G.~Van~Barel, and P.~Hellinckx, ``Automatic reverse engineering of {CAN} bus data using machine learning techniques,'' in \emph{Proc. 3PGCIC '18}, F.~Xhafa, S.~Caball{\'e}, and L.~Barolli, Eds.\hskip 1em plus 0.5em minus 0.4em\relax Cham: Springer International Publishing, 2018, pp. 751--761.

\bibitem{MarkovitzW17}
M.~Markovitz and A.~Wool, ``Field classification, modeling and anomaly detection in unknown {CAN} bus networks,'' \emph{Vehicular Communications}, vol.~9, pp. 43--52, 2017.

\bibitem{opendbc}
{Comma.ai}, ``{OpenDBC},'' \url{https://github.com/commaai/opendbc}, 2017.

\bibitem{VermaIBHMKC22}
\BIBentryALTinterwordspacing
M.~E. Verma, M.~D. Iannacone, R.~A. Bridges, S.~C. Hollifield, P.~Moriano, B.~Kay, and F.~L. Combs, ``Addressing the lack of comparability \& testing in {CAN} intrusion detection research: A comprehensive guide to {CAN} {IDS} data \& introduction of the {ROAD} dataset,'' 2020. [Online]. Available: \url{https://arxiv.org/abs/2012.14600}
\BIBentrySTDinterwordspacing

\bibitem{SongKK16}
H.~M. Song, H.~R. Kim, and H.~K. Kim, ``Intrusion detection system based on the analysis of time intervals of {CAN} messages for in-vehicle network,'' in \emph{Proc. ICOIN '16}, 2016, pp. 63--68.

\bibitem{TomlinsonBSK18}
A.~Tomlinson, J.~Bryans, S.~A. Shaikh, and H.~K. Kalutarage, ``Detection of automotive {CAN} cyber-attacks by identifying packet timing anomalies in time windows,'' in \emph{Proc. IEEE DSN-W '18}, 2018, pp. 231--238.

\bibitem{OlufowobiYZB19}
H.~Olufowobi, C.~Young, J.~Zambreno, and G.~Bloom, ``{SAIDuCANT}: Specification-based automotive intrusion detection using {Controller Area Network (CAN)} timing,'' \emph{IEEE Transactions on Vehicular Technology}, vol.~69, no.~2, pp. 1484--1494, 2020.

\bibitem{YoungOBZ19}
C.~Young, H.~Olufowobi, G.~Bloom, and J.~Zambreno, ``Automotive intrusion detection based on constant {CAN} message frequencies across vehicle driving modes,'' in \emph{Proc. AutoSec '19}, ser. AutoSec '19.\hskip 1em plus 0.5em minus 0.4em\relax New York, NY, USA: Association for Computing Machinery, 2019, pp. 9--14.

\bibitem{KangK16}
M.-J. Kang and J.-W. Kang, ``Intrusion detection system using deep neural network for in-vehicle network security,'' \emph{PLOS ONE}, vol.~11, no.~6, pp. 1--17, 06 2016.

\bibitem{TaylorLJ16}
A.~Taylor, S.~Leblanc, and N.~Japkowicz, ``Anomaly detection in automobile control network data with long short-term memory networks,'' in \emph{Proc. IEEE DSAA '16}, 2016, pp. 130--139.

\bibitem{LongariVZCZ20}
S.~Longari, D.~H. Nova~Valcarcel, M.~Zago, M.~Carminati, and S.~Zanero, ``{CANnolo}: An anomaly detection system based on {LSTM} autoencoders for {Controller Area Network},'' \emph{IEEE Transactions on Network and Service Management}, vol.~18, no.~2, pp. 1913--1924, 2021.

\bibitem{YangTQJWW23}
K.~Yang, X.~Tang, S.~Qiu, S.~Jin, Z.~Wei, and H.~Wang, ``Towards robust decision-making for autonomous driving on highway,'' \emph{IEEE Transactions on Vehicular Technology}, vol. Early access, pp. 1--13, 2023.

\bibitem{obd_ii_pids}
``{OBD-II PIDs} ({Wikipedia}),'' \url{https://en.wikipedia.org/wiki/OBD-II_PIDs}, accessed Jan. 1, 2023.

\bibitem{MuterA11}
M.~M{\"u}ter and N.~Asaj, ``Entropy-based anomaly detection for in-vehicle networks,'' in \emph{Proc. IEEE IV '11}, 2011, pp. 1110--1115.

\bibitem{MarchettiS17}
M.~Marchetti and D.~Stabili, ``Anomaly detection of {CAN} bus messages through analysis of {ID} sequences,'' in \emph{Proc. IEEE IV '17}, 2017, pp. 1577--1583.

\bibitem{TaylorJL15}
A.~Taylor, N.~Japkowicz, and S.~Leblanc, ``Frequency-based anomaly detection for the automotive {CAN} bus,'' in \emph{Proc. WCICSS '15}, 2015, pp. 45--49.

\bibitem{MarchettiSGC16}
M.~Marchetti, D.~Stabili, A.~Guido, and M.~Colajanni, ``Evaluation of anomaly detection for in-vehicle networks through information-theoretic algorithms,'' in \emph{Proc. 2nd IEEE RTSI '16}.\hskip 1em plus 0.5em minus 0.4em\relax IEEE, 2016, pp. 1--6.

\bibitem{StabiliMC17}
D.~Stabili, M.~Marchetti, and M.~Colajanni, ``Detecting attacks to internal vehicle networks through hamming distance,'' in \emph{Proc. AEIT '17}, 2017, pp. 1--6.

\bibitem{KatragaddaDRG20}
S.~Katragadda, P.~J. Darby, A.~Roche, and R.~Gottumukkala, ``Detecting low-rate replay-based injection attacks on in-vehicle networks,'' \emph{IEEE Access}, vol.~8, pp. 54\,979--54\,993, 2020.

\bibitem{SongWK20}
H.~M. Song, J.~Woo, and H.~K. Kim, ``In-vehicle network intrusion detection using deep convolutional neural network,'' \emph{Vehicular Communications}, vol.~21, p. 100198, 2020.

\bibitem{SongKim21}
H.~M. Song and H.~K. Kim, ``Self-supervised anomaly detection for in-vehicle network using noised pseudo normal data,'' \emph{IEEE Transactions on Vehicular Technology}, vol.~70, no.~2, pp. 1098--1108, 2021.

\bibitem{HoangK22}
T.-N. Hoang and D.~Kim, ``Detecting in-vehicle intrusion via semi-supervised learning-based convolutional adversarial autoencoders,'' \emph{Vehicular Communications}, vol.~38, p. 100520, 2022.

\end{thebibliography}
\begin{IEEEbiography}[{\includegraphics[width=1in,height=1.25in,clip,keepaspectratio]{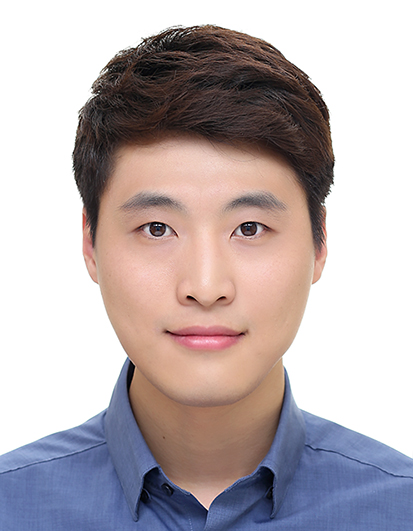}}]{Seonghoon Jeong}
received a Ph.D. degree in information security from the School of Cybersecurity, Korea University, Seoul, Republic of Korea. He is currently a Postdoctoral Researcher with the Institute of Cybersecurity and Privacy, Korea University. His research has focused on in-vehicle network security, including intrusion detection systems for Controller Area Networks and automotive Ethernet.
\end{IEEEbiography}
\vspace{-.5in}
\begin{IEEEbiography}[{\includegraphics[width=1in, height=1.25in, clip, keepaspectratio]{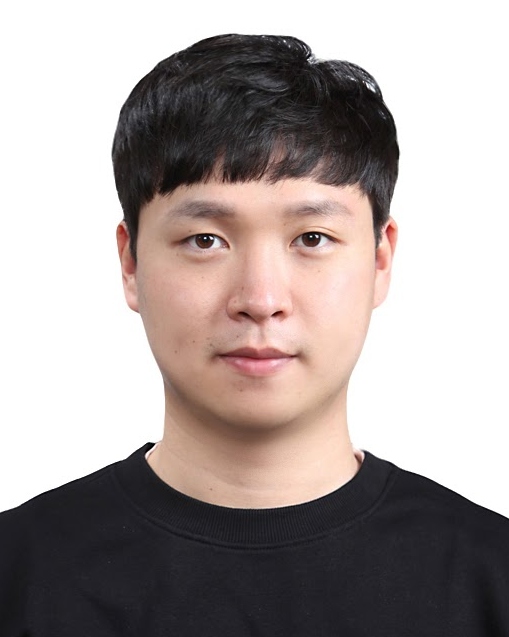}}]{Sangho Lee}
received the B.S. degree in electronic engineering from the Soongsil University in 2018, and the M.S. degree in information security from the School of Cybersecurity, Korea University in 2023. He is a security engineer at Samsung Research of Samsung Electronics since 2018. His research interests include data-driven security, user behavior analysis, and privacy.
\end{IEEEbiography}
\vspace{-.5in}
\begin{IEEEbiography}[{\includegraphics[width=1in, height=1.25in, clip, keepaspectratio]{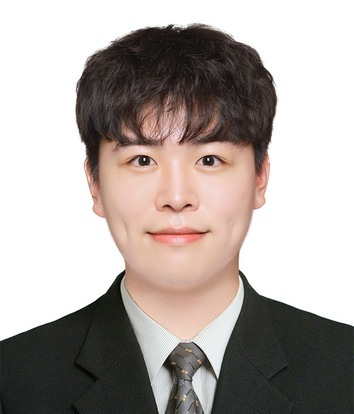}}]{Hwejae Lee}
received a B.S. degree in mechanical engineering from Kyung Hee University in 2020 and is a Ph.D. student in information security at the School of Cybersecurity, Korea University, Seoul, Republic of Korea. His research interests include vehicle security, data-driven security, intrusion detection, machine learning, and deep learning.
\end{IEEEbiography}
\vspace{-.5in}
\begin{IEEEbiography}[{\includegraphics[width=1in, height=1.25in, clip, keepaspectratio]{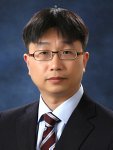}}]{Huy Kang Kim}
received a Ph.D. degree in industrial and system engineering from the Korea Advanced Institute of Science and Technology (KAIST), Republic of Korea. He founded A3 Security Consulting in 1999 and AI Spera, which is a data-driven cyber threat intelligence service company, in 2017. He is a Professor at the School of Cybersecurity, Korea University, Republic of Korea. His recent research has focused on intrusion detection in intelligent transportation systems and in-vehicle networks using machine-learning techniques.
\end{IEEEbiography}
\end{document}